\documentclass[doublespacing]{bmcart}

\usepackage{amsthm,amsmath,graphicx}
\usepackage[utf8]{inputenc} 

\usepackage{amssymb}

\usepackage[right]{lineno}

\usepackage{geometry}
\begin{document}

\begin{frontmatter}

\begin{fmbox}


\title{Accurate chromosome segregation by probabilistic self-organization}


\author[
   addressref={aff1},                   
   corref={aff1},                       
   noteref={n1},                        
   email={y.saka@abdn.ac.uk}   
]{\inits{YS}\fnm{Yasushi} \snm{Saka}}
\author[
   addressref={aff1},
   email={v.c.giuraniuc@abdn.ac.uk}
]{\inits{CVG}\fnm{Claudiu V} \snm{Giuraniuc}}
\author[
   addressref={aff2}, 
   noteref={n1},
   email={h.ohkura@ed.ac.uk}
]{\inits{HO}\fnm{Hiroyuki} \snm{Ohkura}}


\address[id=aff1]{
  \orgname{Institute of Medical Sciences, School of Medical Sciences, University of Aberdeen}, 
  \street{Foresterhill},                     %
  \postcode{AB25 2ZD}                                
  \city{Aberdeen},                              
  \cny{UK}                                    
}
\address[id=aff2]{%
  \orgname{Wellcome Trust Centre for Cell Biology, University of Edinburgh},
  \street{Michael Swann Building, Max Born Crescent},
  \postcode{EH9 3BF}
  \city{Edinburgh},
  \cny{UK}
}


\begin{artnotes}
\note[id=n1]{Correspondence should be addressed to YS or HO.} 
\end{artnotes}

\end{fmbox}



\begin{abstract} 

\parttitle{Background} 
For faithful chromosome segregation during cell division, correct attachments must be established between sister chromosomes and microtubules from opposite spindle poles through kinetochores (chromosome bi-orientation).
Incorrect attachments of kinetochore microtubules (kMTs) lead to chromosome mis-segregation and aneuploidy, which is often associated with developmental abnormalities such as Down syndrome and diseases including cancer. 
The interaction between kinetochores and microtubules is highly dynamic with frequent attachments and detachments.
However, it remains unclear how chromosome bi-orientation is achieved with such accuracy in such a dynamic process.

\parttitle{Results} 
To gain new insight into this essential process, we have developed a simple mathematical model of kinetochore-microtubule interactions during cell division in general, i.e. both mitosis and meiosis. 
Firstly, the model reveals that the balance between attachment and detachment probabilities of kMTs is crucial for correct chromosome bi-orientation. 
With the right balance, incorrect attachments are resolved spontaneously into correct bi-oriented conformations while an imbalance leads to persistent errors. 
In addition, the model explains why errors are more commonly found in the first meiotic division (meiosis I) than in mitosis and how a faulty conformation can evade the spindle assembly checkpoint, which may lead to a chromosome loss.

\parttitle{Conclusions} 
The proposed model, despite its simplicity, helps us understand one of the primary causes of chromosomal instability---aberrant kinetochore-microtubule interactions.
The model reveals that chromosome bi-orientation is a probabilistic self-organization, rather than a sophisticated process of error detection and correction. 
\end{abstract}




%

\end{frontmatter}




\clearpage
\section*{Background}
Accurate segregation of chromosomes during cell division is fundamental to life.  
Errors in this process result in cell death or aneuploidy.  
Chromosome segregation is usually very accurate.
However, mis-segregation occurs at a much higher frequency in cancer cells and oocytes,   
which is a contributing factor to cancer progression \cite{Bakhoum:2009ty} and also a major cause of infertility, miscarriages 
and birth defects such as Down syndrome \cite{Jones:2013qa}. 

The key event for chromosome segregation is the establishment of chromosome bi-orientation, 
in which sister chromatids in mitosis or homologous chromosomes in meiosis attach to the microtubules from opposite spindle poles 
by kinetochores \cite{Tanaka:2010ct}.  
Each kinetochore consists of more than one hundred different proteins assembled on each centromeric DNA sequence, 
many of which are involved in the interaction with microtubules \cite{Cheeseman:2014kq}.  
Chromosome bi-orientation is a very dynamic process with frequent attachments and detachments of microtubules
 \cite{Dietz:1958,Nicklas:1969kq,Kitajima:2011fe,Bakhoum:2012nr}. 
 
For proper segregation of chromosomes, all kinetochores need to attach to spindle microtubules 
while erroneous attachments must be eliminated before anaphase onset. 
It is known that attachment errors are more frequent in meiosis I (especially in oocytes) than 
in mitosis \cite{Dietz:1958,Nicklas:1969kq,Kitajima:2011fe,Jones:2013qa}.
Yet it has not been understood why it is so.
Unattached kinetochores act as signal generators for the spindle assembly checkpoint, 
which delays chromosome segregation until proper bi-orientation is established for all chromosomes \cite{Chen:1996ys}. 
It remains unclear, however, whether improperly attached kMTs are also detected and corrected by the spindle assembly checkpoint 
or by an independent mechanism \cite{Khodjakov:2010uq}.
	
The precise mechanism of chromosome bi-orientation has been under intense investigations.
However, it is not yet possible to observe the dynamics of individual microtubules in vivo in real time.
Mathematical modeling provides a powerful means to study the chromosome bi-orientation process. 
Since the discovery of the dynamic instability of microtubules \cite{Mitchison:1984gf}, a number of theoretical analyses have 
provided important insights into the interaction between microtubules and kinetochores (for example, \cite{Hill:1985cs,Zaytsev07072014}).
The so-called search-and-capture model explains how dynamically unstable microtubules capture chromosomes 
\cite{Kirschner:1986mz,Holy:1994kl,Wollman:2005bk,Gopalakrishnan:2011uo}.
However, the original search-and-capture model did not concern events after capture, in particular, 
erroneous attachments of kMTs and their correction.
To address this, Paul et al. put forward a modified search-and-capture model with explicit correction mechanisms \cite{Paul:2009uq}.
Gay et al. proposed a stochastic model of kinetochore-microtubule attachments in fission yeast mitosis, 
which reproduced correct chromosome bi-orientation and segregation in simulations \cite{Gay:2012dn}. 
In addition to the kinetochore-microtubule interaction, Silkworth \emph{et al.} showed that
timing of centrosome separation also plays a crucial role for accurate chromosome segregation 
\cite{Silkworth:2012kq}; 
using experimental and computational approaches they demonstrated that cells with 
incomplete spindle pole separation have higher rate of kMT attachment errors than those with complete centrosome separation.
Yet, the question remains unanswered as to how the cell can discriminate between correct and incorrect kMT attachments as 
their models assumed an explicit bias based on the discrimination of correct versus incorrect connections.

A major impediment to fully understanding the mechanism of chromosome bi-orientation is the lack of a universal 
theoretical framework that covers the chromosome bi-orientation process during eukaryotic cell divisions in general, including both mitosis and meiosis.
Here we present such a universal model of chromosome bi-orientation, which is simple yet applicable to any eukaryotic cell division.
Firstly, the model reveals that the balance between attachment and detachment probabilities of kMTs is crucial 
for correct chromosome bi-orientation. 
With the right balance, incorrect attachments are resolved spontaneously into correct bi-oriented conformations 
while an imbalance leads to persistent errors. 
Therefore, the superficially complex process, chromosome bi-orientation, is in fact a probabilistic self-organization. 
It implies that the cell does not need to discriminate between correct and incorrect kMT attachments.
Moreover, the model explains why errors are more frequent in meiosis I than in mitosis and 
how a faulty conformation can evade the spindle assembly checkpoint by a gradual increase of the number of kMTs.
Despite its simplicity, the model is consistent with a number of experimental observations and 
provides theoretical insights into the origins of chromosomal instability and aneuploidy.

\section*{Results and discussion}
\subsection*{A probabilistic model of kinetochore-microtubule interaction}
A single kinetochore can bind randomly to microtubules from either left or right pole (Fig. 1A). 
We assume a single kinetochore can accommodate up to $n$ microtubules. 
The process of microtubule attachment/detachment can be represented as a discrete-time 
Markov chain \cite{Norris:1998ez} (Figs. 1B and S1). 
Each pair of sister chromatids in mitosis has two kinetochores ($\mathrm{k_1}$ and $\mathrm{k_2}$ in Fig. 1C).
 In meiosis I, a pair of sister kinecotochores are physically connected side-by-side and act as one \cite{Goldstein:1981kq,Watanabe:2012qf}.
Therefore, in our model, a bivalent (a pair of homologous chromosomes connected by chiasma) also has two kinetochores in meiosis I.
We assume these two kinetochores interact with microtubules independently. 
Hence, the state of the kinetochores is represented as $r_n(i_1,j_1,i_2,j_2)$, 
which can be classified into one of five classes according to the pattern of microtubule attachments (Fig. 1D). 
State transitions occur in a stereotypical manner among these classes irrespective of the value 
of $n\ge2$ (Figs. 1E and S2E; refer to Table 1 for summary of parameters herein).
Notably, the only possible transitions out of class 5 (amphitelic, i.e. correct conformation) is to class 2 
(monotelic) or 4 (merotelic) (red and green arrows in Fig. 1E).
Note also that this transition scheme is similar to the 'kinetic error correction' model (a deterministic ODE model) 
proposed by Mogilner and Craig \cite{Mogilner:2010aa}; their scheme is a limiting case---only two kMT attachments per 
kinetochore are allowed and transitions out of amphitelic states are prohibited.

We assume the association probability is proportional to the available surface area of 
the kinetochore while the dissociation probability is independent, as illustrated below:
\begin{linenomath*}
\begin{eqnarray}
	r_n(i_1,j_1,i_2,j_2)  &\stackrel{ \frac{n-i_1-j_1}{n} p }{\longrightarrow} & r_n(i_1+1,j_1,i_2,j_2),
\label{eqn1}\\
	r_n(i_1,j_1,i_2,j_2) & \stackrel{ i_1  q \;}{\longrightarrow} & r_n(i_1-1,j_1,i_2,j_2),
\label{eqn2}
\end{eqnarray}
\end{linenomath*}
\noindent where $0 \le p\le 1/4$ and $0\le q \le 1/2n$. 
$2\times p$ is the association probability of a single microtubule to a free kinetochore; 
$q$ is the dissociation probability of a single kMT.

 Experimental evidence strongly suggests that tension stabilises the spindle attachment to the kinetochores in amphitelic 
states (class 5) \cite{Nicklas:1994ud,King:2000fk,Dewar:2004pb}.
The stabilisation by tension is brought about by suppression of Aurora B kinase activity towards kinetochore 
substrates \cite{Biggins:2001kq,Tanaka:2002kx,Dewar:2004pb,Cimini:2006pt} as well as by  
mechanical 'catch-bonds' \cite{Akiyoshi:2010fk,Sarangapani:2014gf}. 
We model this stabilisation by scaling the transition probabilities of states in class 5 by detachment 
with the parameter $0\le \beta \le 1$ (Fig. 1F).
This rule also reduces the probability of transitions from class 5 to class 2 states (Fig. 1E, red arrow).
Similarly, the probability of class 5 (amphitelic) to class 4 (merotelic) transitions, 
which occur by attachment of a microtubule but not by detachment (Fig. 1E, green arrow),
scales with $0 \le \alpha \le 1$ (Fig. 1F). 
This is due to the physical constraint imposed in amphitelic states 
in meiosis I \cite{Nicklas:1969kq,Kitajima:2011fe}
or the kinetochore geometry (back-to-back position of sister kinetochores) in mitosis \cite{Tanaka:2010ct}.
In mitosis, $\alpha=0$ for simplicity.
For mitosis we introduce an additional parameter $0\leq \gamma \leq1$ to scale the transition probabilities from class 2 (monotelic) 
to class 3 (syntelic) or 4 (merotelic) (Fig. 1e, blue arrows). 
This is because the biased orientation of sister kinetochores hinders those transitions (Fig. 1F).
Note that when $\alpha=\beta=0$, transitions out of class 5 are effectively blocked; 
hence, this Markov process always ends up in class 5.
For additional details of the model, see Supplementary Information (SI) Text.
This simple model, which has only six parameters and is exactly solvable, 
provides a number of analytical insights into how correct chromosome bi-orientation is achieved.

\subsection*{Dynamics of chromosome bi-orientation process}
The model predicts how long it takes to reach class 5 (amphitelic) from class 1 (free), i.e. 
mean first passage time \cite{bertsekas2008introduction} (see SI Text).
For a given value of $q$, the mean first passage time 
(which is independent of $\alpha$ and $\beta$ because they only affect transitions out of class 5) is shortest 
when $p$ is roughly equal to $q$ (Figs. 2A and S3A-D).
Thus, the relative dissociation rate ($q/p$ ratio) of kMTs needs to be 
balanced for efficient chromosome bi-orientation.

The model also predicts the dynamics of the system (Fig. 2B-D for meiosis I and E-G for mitosis).
Note that the $q/p$ ratio dictates the dynamics of the Markov chain (Fig. S5).
For both mitosis and meiosis in an ideal condition ($p=q=0.05, \alpha=\beta=0$; Fig. 2B, E), the probability of class 5 
steadily increases, asymptotically reaching 1.
Notably, in meiosis I, class 4 (merotelic), and class 3 (syntelic) to a lesser extent, become transiently prominent (Fig. 2B).
Merotelic attachments are indeed frequently observed in prophase to prometaphase of meiosis I 
in mouse oocytes \cite{Kitajima:2011fe}.
By contrast, in mitosis, class 2 (monotelic) becomes predominant before replaced by class 5, 
although minor fractions of class 3 and 4 also appear briefly (Fig. 2E). 
Together, it explains why meiosis I is more error-prone than mitosis;
it is attributed to the parameter $\gamma$\,---\,
the back-to-back conformation of sister kinetochores, which biases the kinetochore orientation.

If there is no bias in meiosis I (random condition; $\alpha=\beta=1$; Fig. 2C, see also Fig. S4), 
the probability of class 5 stays low while that of class 4 (merotelic) reaches nearly one half at steady states.
This is because class 4 is by far the largest among the five classes (Fig S2A and S2B).
In mitosis, when the spindle tension is lacking ($\beta=1$; Fig. 2F), the model predicts high probability of 
errors, mainly monotelic (class 2) states,  as well as the correct amphitelic (class 5) ones at steady states.
When kinetochore-microtubule attachment is stabilised by reducing $q$, merotelic errors (class 4)
persists in both meiosis and mitosis (Fig. 2D, G). 
Class 5 will eventually replace class 4 but only very slowly;
in meiosis I with $p=0.05, \alpha=\beta=0$, the mean first passage times to class 5 are 
$\sim1631$ for $q=0.01$ versus 
$\sim47$ for $q=0.05$.

A number of studies demonstrated that experimental manipulations of 
kinetochore-microtubule interactions lead to accumulation of incorrect spindle 
attachments (class 1-4) and aneuploidy \cite{Bakhoum:2012nr}.
Lack of tension (i.e. $\beta=1$) makes amphitelic states (class 5) unstable \cite{Nicklas:1994ud,King:2000fk,Dewar:2004pb}.
Conversely, inhibition or depletion of Aurora B kinase, which over-stabilizes kMT attachments (by reducing $q$),
caused errors in chromosome alignment and segregation 
\cite{Hauf:2003ys,Dewar:2004pb,Cimini:2006pt,Kelly:2009uq,Kitajima:2011fe}.
These observations are consistent with our model predictions in which imbalance of 
the $q/p$ ratio causes persistent errors in kMT attachments (Fig. 2).

\subsection*{Probability distribution of the number of kMTs over time}
Next, we calculated the probability distribution of the number of kMTs over time in different conditions (Figs. 3A-C and S6 for meiosis I; Fig. S7 for mitosis).
We found qualitatively similar kMT distributions in mitosis and meiosis I, 
 except the difference in the predicted phenotype in various conditions (Fig. 2B-G).
The model predicts that in 'normal' conditions ($p=q=0.05,\alpha=\beta=0$) the number of kMTs increases 
steadily in class 5 while it remains low in the other classes 
as their total probability diminishes (Figs. 3A and S7A).
This is in agreement with experimental evidence suggesting the gradual 
increase of kMTs during prometaphase to metaphase in mitosis \cite{McEwen:1997zt} and in meiosis I \cite{King:2000fk}.
With smaller $q$, the number of kMTs increases not only in class 5 but also in class 4 
(Figs. 3B and S7B).
This explains why errors persists in this condition.
Note that when $\beta=0$, the number of kMTs approaches $n$.
Increasing number of kMTs may also switch off the spindle assembly checkpoint in merotelic (class 4) states over time.

These model predictions on the probability distribution of the number of kMTs have an important implication 
in the regulation of spindle assembly checkpoint.
Experimental evidence suggests that intrakinetochore stretching (or kinetochore deformation), 
which is brought about by kMT attachments, has a role in relieving the spindle assembly checkpoint 
\cite{Uchida:2009qy,Maresca:2009jk,Nannas:2014qy}.
Therefore the predicted gradual increase of kMTs in amphitelic (class 5) states (Figs. 3A and S7A) may 
switch off the spindle assembly checkpoint progressively.
The same argument applies to merotelic (class 4) states, the probability of which increases 
when $q/p$ ratio is small (Fig. 2D, G); 
stabilisation of kMTs (Figs. 3B and S7B) may also inactivate the spindle assembly checkpoint in merotelic states over time.
This provides the explanation as to why merotelic orientation evades the spindle assembly checkpoint 
\cite{Cimini:2001fk}, leading to aneuploidy.
Intrakinetochore stretching by kMT attachment, however, does not allow the cell to discriminate between correct (amphitelic; class 5) 
versus incorrect (non-amphitelic; class 1-4) kMT attachments \cite{Khodjakov:2010uq}\,---\,the cell does not need to do so
because chromosome bi-orientation occurs by probabilistic self-organisation as our model indicates.

We also examined how kMT number changes in amphitelic states under low spindle tension 
($\beta=1$; Figs. 3C and S7C). 
Regardless of the classes, the distribution of kMT number remains low, 
which makes the transition of the process from one class to another more frequent. 
Similar probability distributions of kMT number in meiosis I were obtained when $\alpha=\beta=1$ 
(Fig. S6A) and $\alpha=1, \beta=0$ (Fig. S6B).

The exact probability distribution of kMT number at steady states can be derived in the special case when 
$\alpha=\beta=\gamma=1$: 
its mean is $\bar{N}=n\rho/(n+\rho)$ where $\rho=2 p/q$ ($\bar{N}=5/3$ for $p=q, n=10$).
We also obtained an analytical approximation of the kMT number distribution in class 5 when $\alpha=0$:
\begin{linenomath*}
\begin{equation}
\bar{N}_5=\frac{\bar{\rho}  \left(\frac{\bar{\rho} }{n}+2\right)^{n-1}}{\left(\frac{\bar{\rho} }{n}+2\right)^n-2^n},
\end{equation}
\end{linenomath*}
\noindent where $\bar{\rho}=\rho/\beta=2p/(\beta q)$ (Figs. 3D and S8A, B).  
This formula is valid for both mitosis and meiosis and provides an analytical explanation as to how tension ($\beta$) alters
 the stability of kMTs by modulating the $q/p$ ratio. 

\subsection*{Dynamics of multiple chromosomes}
The above results concern the behaviour of a single pair of homologous chromosomes.
It is natural to ask how multiple pairs in the cell are bi-oriented simultaneously---we call this event 'synchrony' to distinguish it from anaphase onset.
We assumed the system consists of $k$ independent Markov processes.
Let $\theta_t$ be the probability of a process being in class 5 (amphitelic) at time $T=t$, then the probability
 of synchrony at $T=t$ is $\theta_t{}^k$ (see SI Text).

The timing of synchrony delays as $k$ increases (Figs. 4A and S3E, solid lines). 
If the balance of $q/p$ ratio is broken by reducing $q$ (Figs. 4A  and S3E, dashed lines), 
the timing of synchrony is delayed further (see also Fig. S9).
The probabilities of synchrony, however, eventually approach 1 in all of these conditions with $\beta=0$.
This implies that delaying the onset of anaphase could reduce the chromosome 
mal-orientation and mis-segregation. 
Consistently, Cimini et al. showed that prolonging metaphase significantly reduced lagging chromosomes 
in anaphase (indicating incorrect kMT attachments) in mitosis \cite{Cimini:2003uq}.
 
We next examined the contribution of $\alpha$ and $\beta$ to the establishment of synchrony. 
Fig. 4B shows the steady-state probability of synchrony in meiosis I as a contour plot. 
It indicates that, to achieve a synchrony reliably at 
steady states, $\alpha$ and $\beta$ have to be relatively small. 
It is conceivable that, to progress into anaphase, synchrony has to be maintained for a 
sufficient time to relieve the spindle assembly checkpoint \cite{Khodjakov:2010uq}. 
 Fig. 4C depicts the half-life of synchrony in meiosis I as a contour plot (see also Fig. S3F for mitosis). 
 The half-life increases steeply towards the small values of $\alpha$ and $\beta$.
These data suggest that $\alpha$ and $\beta$ need to be tightly regulated for efficient 
chromosome bi-orientation and segregation accuracy.
 
 \subsection*{Error correction of kMT attachments in meiosis I}
Finally, we asked how many rounds of error correction of kMT attachments occur in meiosis I
before the establishment of correct bi-orientation (see SI Text for methods).
We calculated the number of bi-orientation attempts per bivalent, i.e. the mean number of 
transitions from class 2 or 4 to class 5 before the kinetochore is fully occupied
($r_n(n,0,0,n)$ and $r_n(0,n,n,0)$ when $\beta=0$) (Fig. 4D).  
It suggests that the larger $\alpha$, the more bi-orientation attempts. 
We also found the number of bi-orientation attempts decreases as $q$ (detachment probability) reduces (Fig. 4D, see also Fig. S10). 
Consistent with this, Kitajima et al. observed the number of attempts reduced from $\sim$
\hspace{-.12ex}$3$ in untreated mouse oocytes to just one on average in those treated with 
Hesperadin, an Aurora B kinase inhibitor \cite{Kitajima:2011fe}.

\section*{Conclusions}

Our simple discrete-time Markov chain model captures the prominent features of chromosome bi-orientation process.
It provides a unified account of two modes of divisions, mitosis and meiosis I, under a single theoretical framework;
 the model reveals where the differences in the bi-orientation process come from. 
It explains why errors are very frequent in the first meiotic division, which are major causes of infertility, 
miscarriages and birth defects in humans.

One of our key findings in this study is that the system dynamics (including the type and frequency of transient kMT attachment errors)
 is dictated by the $q/p$ ratio (relative detachment rate) of kMTs. 
An imbalance of $q/p$ ratio causes persistent attachment errors leading to chromosome mis-segregations. 
The gradual increase of kMTs may help turn off the spindle assembly checkpoint in normal conditions but can promote a faulty conformation
 (merotelic attachments) to evade the checkpoint.

In summary, our study revealed that the chromosome bi-orientation is a probabilistic self-organization, 
rather than a sophisticated process of error detection and correction. 
Although our model omits many potentially important factors for chromosome bi-orientation, 
such as the spatial arrangement of centrosomes, 
it allowed us to examine analytically all possible outcomes with different parameters (i.e. the whole parameter space), 
revealing what is fundamental to accurate chromosome segregation. 
The proposed model, which is based on a firm mathematical foundation, gives valuable insights that help us understand one of the primary causes of 
chromosomal instability---aberrant kMT dynamics.

\section*{Methods}
The model and its analysis are explained in detail in SI Text (Additional File 1). 
The analysis of discrete-time Markov chains were performed according to 
\cite{Norris:1998ez,bertsekas2008introduction,Kemeny:1960gs}.
We used \textit{Mathematica}\textsuperscript{$\circledR$} (version 10, Wolfram Research) for the analysis of the model, 
with a standard laptop (or desktop) computer.
The \textit{Mathematica} codes used in this study are provided in Additional File 2.



\section*{Competing interests}
  The authors declare that they have no competing interests.

\section*{Author's contributions}
   YS and HO designed the model; YS wrote the computer codes and analysed the model; CVG analysed the model; and all authors wrote the paper.

\section*{Acknowledgements}
  We thank G. Bewick, C. Grebogi, S. Hoppler, A. Lorenz, C. McCaig, F. Perez-Reche, R. Sekido, M. Thiel and E. Ullner for helpful discussions and critical reading of the manuscript. YS and CG are supported by Scottish Universities Life Sciences Alliance (SULSA) and HO by Wellcome Trust (grant number 098030 and 092076). 
  

\bibliographystyle{bmc-mathphys} 
\bibliography{saka_refs} 



\newpage

\section*{Figures}
  \begin{figure}[h!]
  \includegraphics[width=10cm]{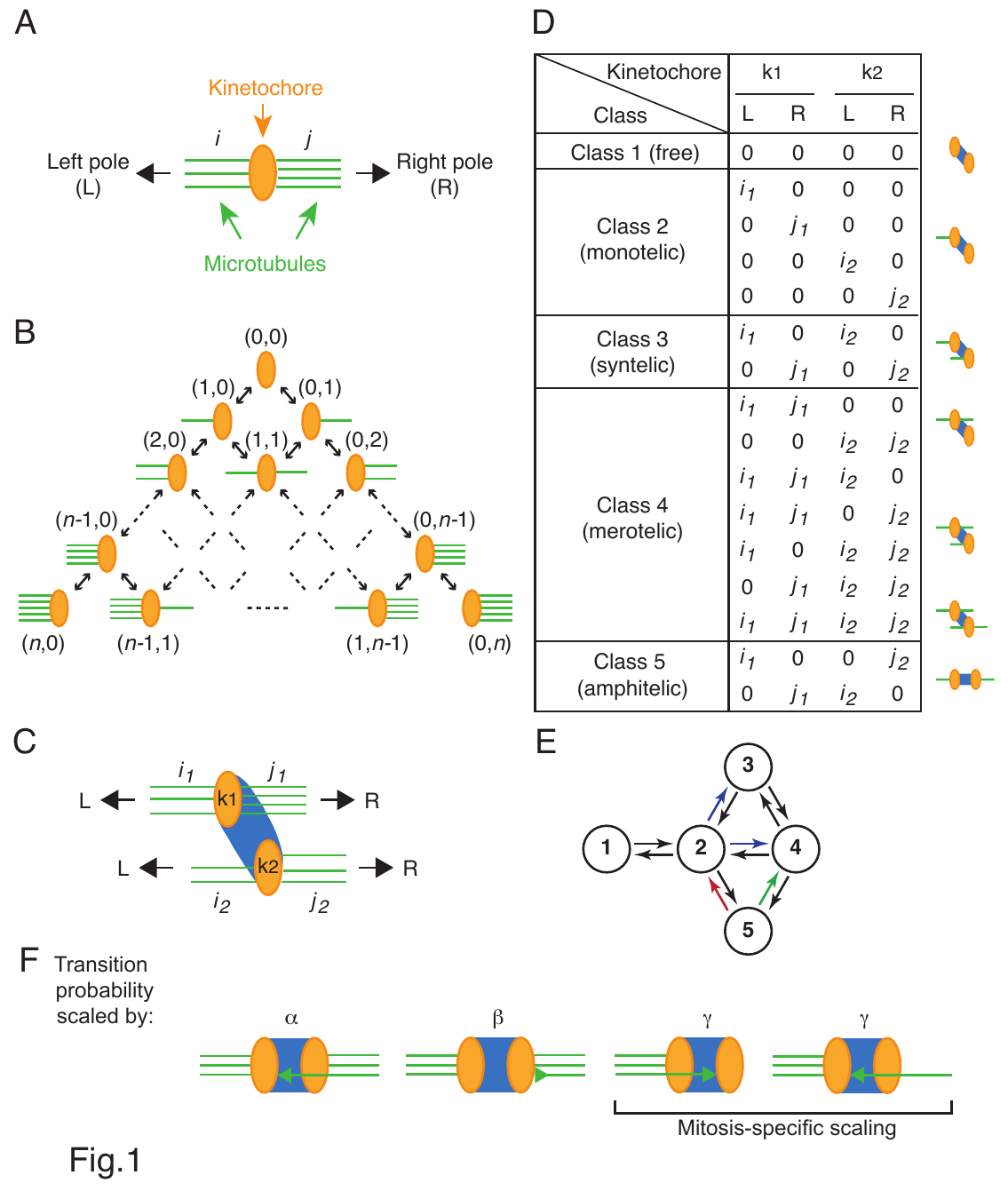}
  \caption{\csentence{A discrete-time Markov chain model of kMT dynamics.}
 (A) Schematic diagram of the interaction between a kinetochore (orange) and microtubules (green) from either left (L) or
right (R) pole. $i$ and $j$ indicate the number of kMTs. (B) Kinetochore-microtubule interactions as a
Markov chain. The maximal number of kMTs per kinetochore is $n$. (C) Schematic diagram of kMT
dynamics during cell division. A pair of kinetochores (k1 and k2) are connected by bivalent
chromatids in meiosis I or centromere chromatins (blue). (D) States of kinetochore-microtubule
complex are defined with $r_n(i_1, j_1, i_2, j_2)$. Every state can be classified into one of five classes in
the table. Schematic diagrams of each class are shown on the right. (E) Transition diagram among
classes. A subset of states in the Markov chain categorised in (D) can move from one class to
another according to this diagram. To increase the probability of class 5 states, transitions out of
class 5 (red and green arrows) must be reduced, the probabilities of which are scaled with
parameters $\alpha$ (for green arrow) and $\beta$ (for red arrow) in the model. In mitosis, transitions from
class 2 to class 3 or 4 are scaled with $\gamma$ (blue arrows). (F) Schematic diagram of the scaling by
parameters $\alpha$, $\beta$ and $\gamma$. Probabilities of state transition by attachment or detachment (arrowheads)
are scaled by the indicated parameters.}
      \end{figure}

\newpage

\begin{figure}[h!]
  \includegraphics[width=10cm]{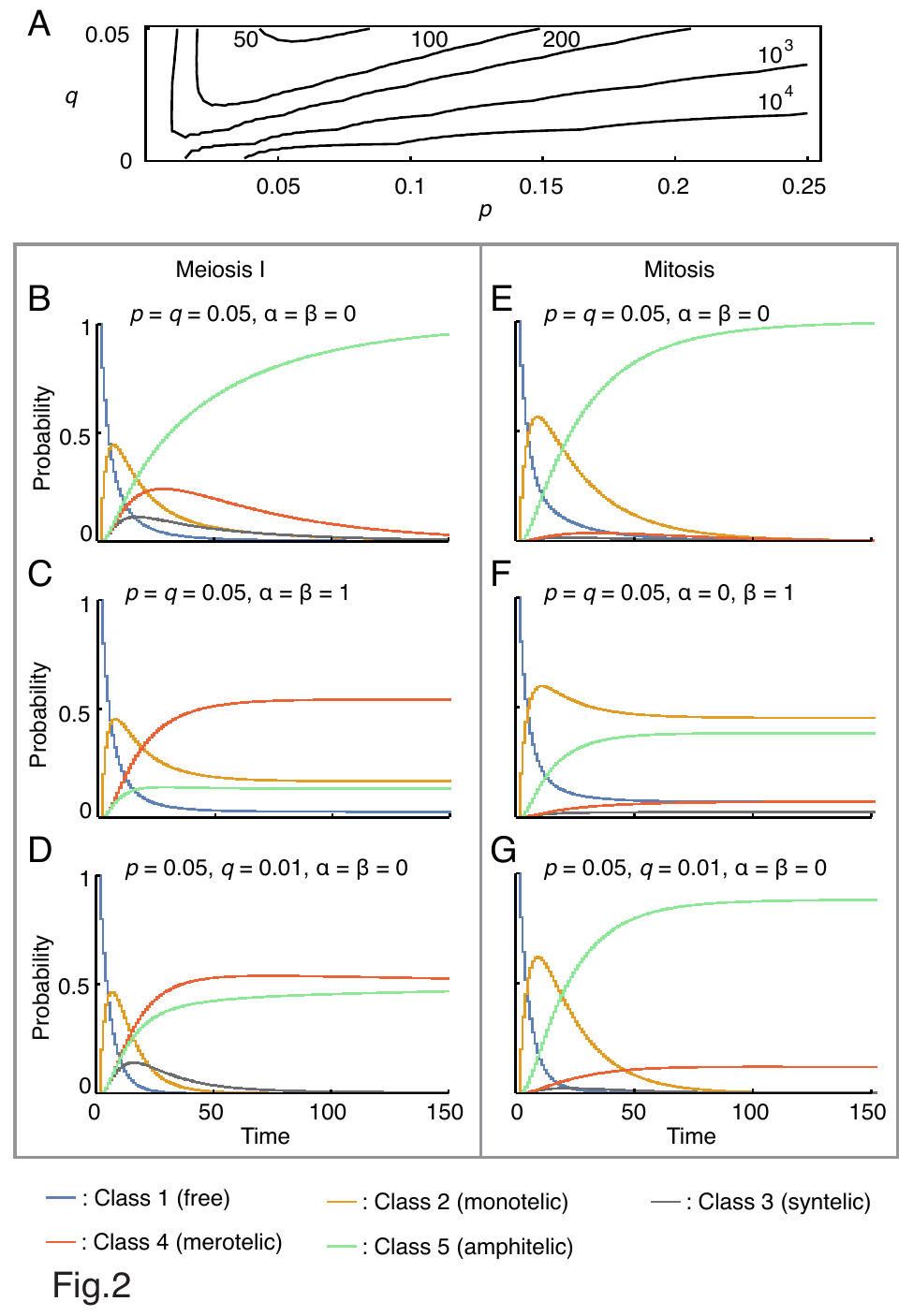}
  \caption{\csentence{Dynamics of kinetochore-microtubule interaction.}
(A) Contour plot of mean first passage time to class 5 starting from class 1 in meiosis I. (B-G) Probabilities of each class over
time for meiosis I (B-D) and mitosis (E-G). $n = 10$ for all panels. $\gamma = 1$ for meiosis I and $\gamma = 0.1$ for
mitosis. Other parameters are as indicated for each panel. (B, E) An 'ideal' condition. The
probability of class 5 approaches 1. (C, F) A 'random' condition with no bias towards class 5. The
probability of class 4 (merotelic) becomes predominant in meiosis I (C) while class 2 (monotelic)
is as prevalent as class 5 (amphitelic) in mitosis (F). Note that class 3 and class 5 have identical
probabilities by symmetry in (C). (D, G) A condition in which $q/p$ ratio is low. 
Class 4 persists both in meiosis I and in mitosis.}
      \end{figure}

\newpage

\begin{figure}[h!]
  \includegraphics[width=10cm]{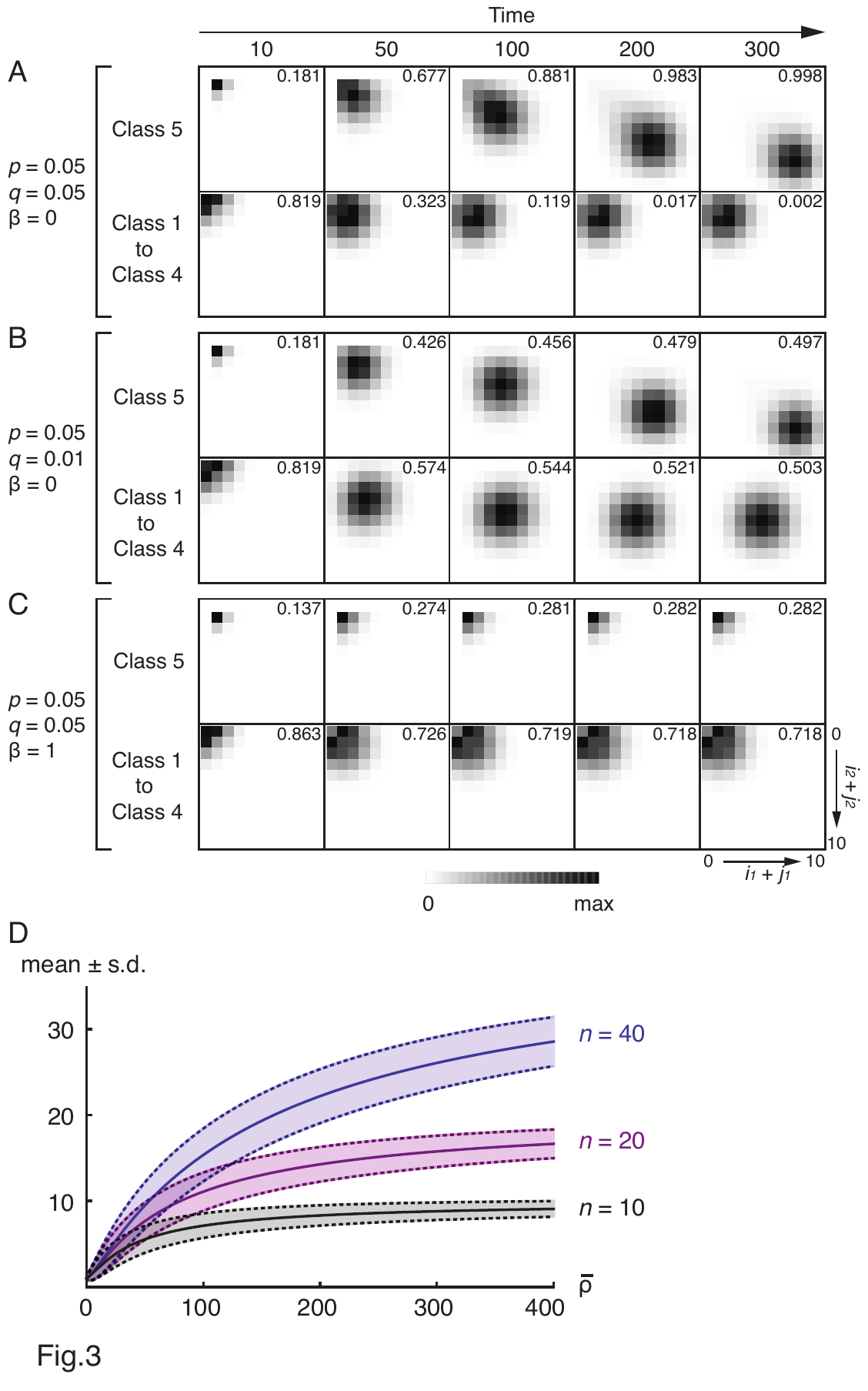}
  \caption{\csentence{Probability distribution of the number of kMTs over time.}
(A-C) Probability density plots of the number of kMTs in meiosis I in 2D ($i_1 + j_1$ vs. $i_2 + j_2$; see Fig. 1C) at the indicated time points. 
Parameters are indicated on the left. $\alpha = 0$, $n = 10$ for all panels. 
Probabilities are decomposed into class 5 and the rest (class 1 to 4) at each time point. 
Total probabilities are indicated on each panel. 
The densities are scaled from 0 to the maximal for each panel. 
(D) Mean number of microtubules ($\pm s.d.$) attached to a kinetochore derived by the approximation formulae
(Eqs. (3) and (10)). Plots for $n = 10$, $20$ and $40$ are shown. For details, see SI Text.}
      \end{figure}

\newpage

\begin{figure}[h!]
  \includegraphics[width=10cm]{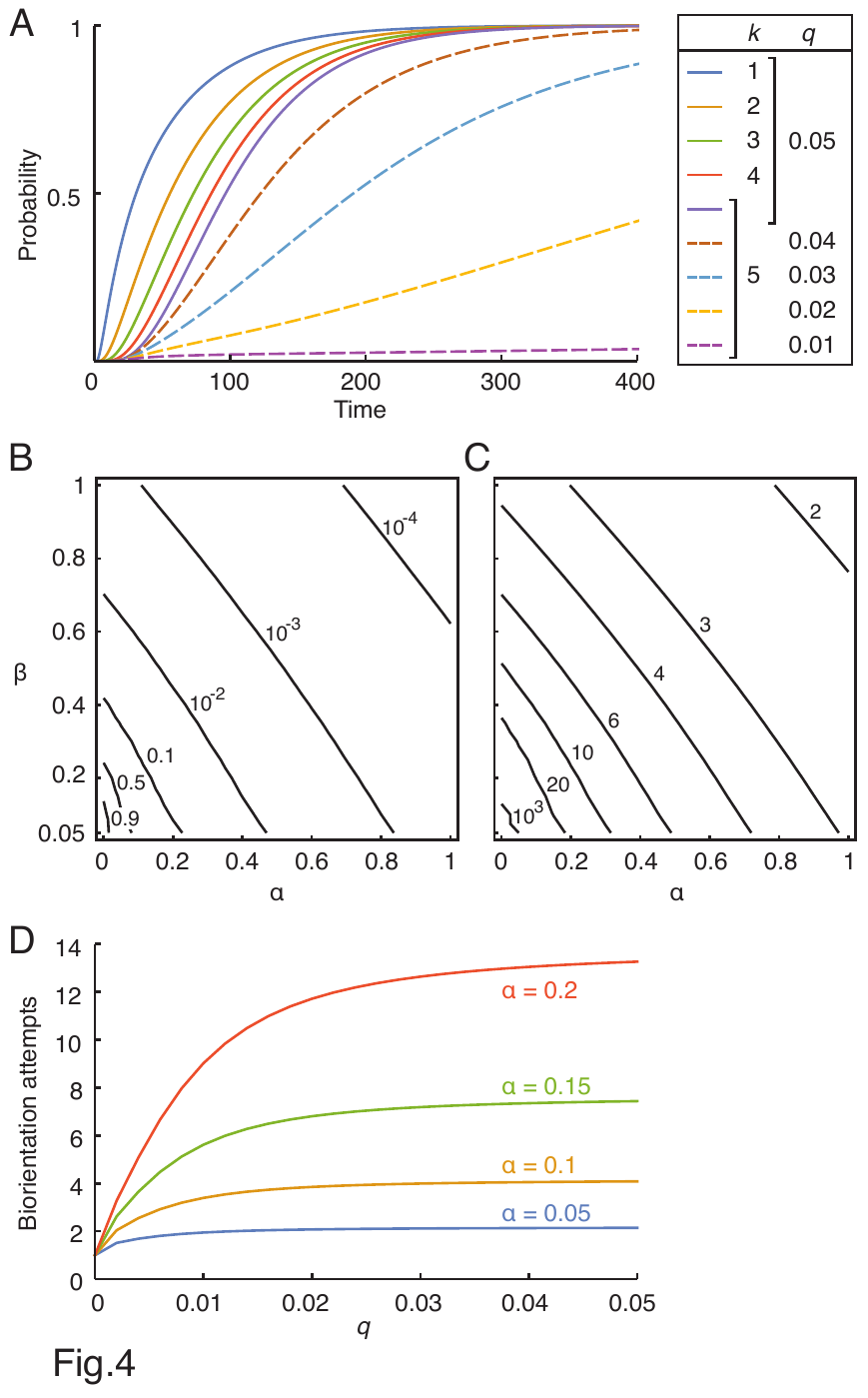}
  \caption{\csentence{Dynamics of multiple chromosomes in meiosis I.}
(A) Probabilities of synchrony over time. $k = {\text{number of chromosomes}}$; $p = 0.05$, $\alpha = \beta = 0$. 
(B) Contour plot of probability of synchrony at steady states. 
(C) Contour plot of half-life of synchrony at steady states. In (B) and
(C), $p = q = 0.05$, $k = 5$. (D) Number of biorientation attempts before absorption. $p = 0.05$, $\beta = 0$.
$n = 10$ for all panels.}
      \end{figure}


\newpage

\section*{Tables}
\begin{table}[h!]
\caption{Model parameters.}
      \begin{tabular}{clcl}
        \hline
           & Parameter for  &Range of value   & Biological meaning\\ \hline
        $n$ & Maximal number of kMTs & $2\le n$ & Maximal number of kMTs that can be accommodated on a single\\ 
        &per kinetochore&& kinetochore. $n$ is proportional to the size of a kinetochore.\\
        $p$ & Association probability &  $0\le p \le 1/4$ & $2\times p$ is the association probability of a single microtubule to a free\\
        &&& kinetochore in each discrete time step. Upper limit of $p$ is 1/4 \\
        &&& because total probability $\le 1$.\\
        $q$ & Dissociation probability & $0\le q \le 1/2n$ & Dissociation probability of a single kMT in each discrete time step.\\
        $\alpha$ & Scaling factor of $p$ & $0\le \alpha \le 1$   & Scaling applies to transitions from amphitelic (class 5) to merotelic \\
        &&&(class 4) states; reflecting the physical constraint imposed in  \\
        &&&amphitelic states (meiosis I) or the back-to-back position of sister \\
        &&&kinetochores (mitosis). $\alpha=0$ in mitosis for simplicity.\\
        $\beta$ & Scaling factor of $q$ & $0\le \beta \le 1$ & Scaling applies to transitions in/from amphitelic states (class 5); \\
        &&& reflecting the kMT stabilization by tension. \\
        $\gamma$ & Scaling factor of $p$ &$0\leq \gamma \leq1$ & Scaling applies to transitions from monotelic (class 2) to syntelic \\
        &&&(class 3) or merotelic (class 4) states in mitosis; reflecting the \\
        &&&biased orientation of sister kinetochores in monotelic states. \\\hline
      \end{tabular}
\end{table}




\newpage
\section*{Supplementary Information}

\section{A basic Markov chain model of kinetochore-microtubule interactions (Model I)}

The interaction of a single kinetochore with microtubules is modeled as a birth/death (discrete-time) Markov process. 
First, we consider a kinetochore that can bind up to \textit{n} microtubules. 
The possible states are \(M=\{0,1,2,\text{...},n\}\). 
Transition probability from state \textit{ i} to \textit{ j} is \(p_{i, j}= P\left(X_{t+1}= j \left| X_t\right.= i \right), \text{  }i, j\in M\), 
where \(X_t\) is the state at time\textit{ t}.
As stated in the main text, we assume the association probability is proportional to the surface area of a kinetochore available 
for microtubule attachment. 
Therefore, the association (birth) probability is 
$p_{k, k+1}=(n-k)\,b/n,\;0\leq  k\leq n-1$ 
 where \(b\) is the association probability of a single microtubule to a free kinetochore. 
 The dissociation (death) probability for state \(k\) is $p_{k, k-1}= k\,d,\;1\leq  k\leq n$, 
 because each microtubule bound to a kinetochore (\(k\) microtubules in total) has the same dissociation probability \(d\). 
 The self-transition probability is
$p_{k, k}=1-(n-k)\,b/n-k \,d,\;0\leq  k\leq  n$. 
As an example, consider a kinetochore that can bind up to 2 microtubules. 
The possible states are \(\{0,1,2\}\).
The transition probability matrix is 
\begin{linenomath*}
\begin{equation*}
 R_3=\left(
\begin{array}{ccc}
 p_{0, 0} & p_{0, 1} & p_{0, 2} \\
 p_{1, 0} & p_{1, 1} & p_{1, 2} \\
 p_{2, 0} & p_{2, 1} & p_{2, 2} \\
\end{array}
\right) 
 =\left(
\begin{array}{ccc}
 1-b & b & 0 \\
 d & 1-\frac{1}{2}b-d & \frac{1}{2}b \\
 0 & 2d & 1-2d \\
\end{array}
\right). 
\end{equation*}
\end{linenomath*}
 
\noindent  Fig. S1A shows a diagram of this Markov chain. 
This model is a variation on the M/M/s queue \cite{Norris:1998ez}. 
The Markov chain consists of a single aperiodic recurrent class. Let \(u_k\) be the steady-state probabilities of state \(k\) 
and \(U\) be the row vector of \(u_k, k=0, 1,2,\text{...},n\).
Applying the steady-state convergence theorem, then $U=U R_n$. 
This is equivalent to a local balance equation: 
$ (1-k/n)\,b \,u_k=(k+1)d \,u_{k+1}. $ Let \(\rho =b/d\) 
then,
 \begin{linenomath*}
\setcounter{equation}{3}
\begin{equation}
 (1-k/n)\rho  u_k=(k+1) u_{k+1},k=0,1,\text{...},n-1. \\
 \label{balance}
\end{equation}
 \end{linenomath*}
\noindent The normalization equation (the sum of all probabilities equals to 1) is
 \begin{linenomath*}
\begin{equation}
\sum\limits_{k=0} ^n u_k =1.
\label{norm}
\end{equation}
 \end{linenomath*}
\noindent Eqs. (\ref{balance}) and (\ref{norm}) yield a unique solution:
 \begin{linenomath*}
\begin{equation}
u_k =\binom{n}{k}(\rho /n)^k(1+\rho /n)^{-n}, \;k=0,1,2,\text{...},n, \\
\label{sol1}
\end{equation}
 \end{linenomath*}
\noindent  where $\binom{n}{k}$ is the binomial coefficient 
 ($\texttt{"}$\textit{n} choose \textit{k}$\texttt{"}$). 
 When \(n\) is large, \(u_k\) approaches the Poisson distribution \(e^{-\rho } \left.\rho ^k\right/k!\). 
The mean and variance of \(u_k\), derived from Eq. (\ref{sol1}), are \(n \rho /(n+\rho )\) and
\(n^2 \rho \left/(n+\rho )^2\right.\), respectively. 
Fig. S1B shows an example of the probability distribution of \(u_k\) (the number of attached
microtubules) for \(n=20\). 
As illustrated in this example, the stability of kinetochore-microtubule interaction can be controlled by $\rho $ (i.e.
\(d/b\) ratio) alone. 

\section{The extended model of kinetochore-microtubule interactions (Model II)}
Now we consider the interaction of a kinetochore with bipolar spindles (Fig. 1B in the main text). There are \((n+1)(n+2)/2\) possible states, e.g.
6 possible states for \(n=2\). We assign a unique index number to each state denoted as \(s_n(i,j)\):
 \begin{linenomath*}
\begin{equation} 
 s_n(i,j)\longmapsto i+1+\frac{(i+j+1)(i+j)}{2},  \text{  } 0\leq i+j\leq  n. \\
\label{index}
\end{equation}
 \end{linenomath*}
\noindent We use these indices to construct the probability transition matrix in \textit{Mathematica} codes.
Using the same argument for model I, the state transition probabilities are
 \begin{linenomath*}
\begin{eqnarray}
s_n(i,j) &\overset {\frac{n-i-j}{n}p}{\longrightarrow} & s_n(i+1,j), \nonumber\\
s_n(i,j) &\overset {\frac{n-i-j}{n}p}{\longrightarrow} & s_n(i,j+1),\label{rule2}\\
s_n(i,j) &\overset {i \,q}{\longrightarrow} & s_n(i-1,j),\nonumber\\
s_n(i,j) &\overset {j \,q}{\longrightarrow} & s_n(i,j-1), \nonumber
\end{eqnarray}
 \end{linenomath*}
\noindent where \textit{p}, \textit{q} are parameters with \(0\leq  p\leq  1/2,  \text{  } 0\leq  q\leq  1/n\). 
For example, the transition matrix \(P_n\) with \(n=2\) is 
 \begin{linenomath*}
\begin{equation*}
P_2=\left(
\begin{array}{cccccc}
 1-2 p & p & p & 0 & 0 & 0 \\
 q & 1-p-q & 0 & \frac{p}{2} & \frac{p}{2} & 0 \\
 q & 0 & 1-p-q & 0 & \frac{p}{2} & \frac{p}{2} \\
 0 & 2 q & 0 & 1-2 q & 0 & 0 \\
 0 & q & q & 0 & 1-2 q & 0 \\
 0 & 0 & 2 q & 0 & 0 & 1-2 q \\
\end{array}
\right). \\
\end{equation*}
 \end{linenomath*}
 
\noindent Model II is fundamentally the same as Model I: \(b\) in Model I is equal to the combined probability of 
a microtubule binding to a free kinetochore (\(=2\times p\)) in Model II. 
Dissociation probability of a single kMT is the same (\(d=q\)). Hence \(\rho =2p/q\).

\section{The model of kinetochore-microtubule interactions in meiosis and mitosis (Model III)}

The model of kinetochore-microtubule interactions in meiosis and mitosis is built from Model II, 
which we call Model III. 
This model describes the state of a pair of kinetochores physically connected 
by a centromere chromatin (in mitosis and meiosis II) or a bivalent (in meiosis I) of homologous 
chromosomes, which is defined by $r_n\left(i_1,j_1,i_2,j_2\right)$. 
Note that $0 \le p\le 1/4$ and $0\le q \le 1/2n$ in Model III because the total transition probability from 
a given state including self-transition is 1. 
Also note that there is no direct transition from class 1 (free) to class 5 (amphitelic, i.e. correct conformation).

As briefly mentioned in the main text, spindle tension stabilises the kMT attachments in amphitelic 
states (class 5), which is represented by the scaling with the parameter $\beta$. 
This applies to both mitosis and meiosis. 
The scaling with the parameter $\beta$ is exemplified by
 \begin{linenomath*}
\begin{equation*}
r_n(i_1,0,0,j_2)  \stackrel{ i_1 \beta \,q\;\;\;}{\longrightarrow}  r_n(i_1-1,0,0,j_2).
\label{eqn3}
\end{equation*}
 \end{linenomath*}
\noindent This rule also reduces the probability of transitions from class 5 to class 2 states (red arrow in Fig. 1E; 
with $i_1=1$ in the above example).

The scaling of the probability of class 5 (amphitelic) to class 4 (merotelic) transitions with the parameter $\alpha$ is based on 
the experimental evidences.
In amphitelic states in mitosis, the kinetochore geometry in mitotic chromosomes prevents each sister
 kinetochore from interacting with the microtubules from the opposite pole \cite{Tanaka:2010ct}. 
 Therefore class 5 (amphitelic) to class 4 (merotelic) transitions are effectively eliminated in mitosis, i.e. $\alpha=0$.
In meiosis I, Nicklas suggested that the stability of amphitelic conformation is also gained by the aligned position of kinetochores 
with the pole-to-pole axis, with each kinetochore pointing at a pole \cite{Nicklas:1969kq}. 
A recent study of meiosis I in mouse oocyte indeed revealed the restricted movement of kinetochores in amphitelic states
(see supplemental movies in Kitajima et al \cite{Kitajima:2011fe}). 
The scaling with the parameter $\alpha$ is exemplified by
\begin{linenomath*}
\begin{equation*}
r_n(i_1,0,0,j_2)  \stackrel{ \frac{n-i_1}{n} \alpha \,p}{\longrightarrow}  r_n(i_1,1,0,j_2).
\label{eqn4}
\end{equation*}
\end{linenomath*}
With a similar reason transitions from class 2 (monotelic) to class 3 (syntelic) or 4 (merotelic) are reduced in mitosis 
because the attached sister kinetochore are facing towards the pole from which the kMT emanates, while the other unattached 
 sister kinetochore are facing the opposite pole. 
 Thus, these transitions (blue arrows in Fig. 1E) are scaled by the parameter $0\leq\gamma\leq1$. 
 For example,
 \begin{linenomath*}
 \begin{eqnarray*}
r_n(i_1,0,0,0)  &\stackrel{ \gamma \,p}{\longrightarrow} & r_n(i_1,0,1,0),
\label{eqn5}\\
r_n(i_1,0,0,0)  &\stackrel{ \frac{n-i_1}{n}\gamma \,p}{\longrightarrow}&  r_n(i_1,1,0,0).
\label{eqn6}
\end{eqnarray*}
\end{linenomath*}
\noindent These scaling of the transitions by $\gamma$ are unique to mitosis (and meiosis II); 
for meiosis I, $\gamma=1$.

With sufficiently small $\alpha$ and $\beta$, class 5 becomes stable; when $\alpha=\beta=0$, transitions out of class 5 are not possible. 
That means class 5 is an absorbing class in the Markov chain. 
This bias towards class 5 underpins the probabilistic self-organisation of the system.
By contrast, when $\alpha=\beta=1$ there is no bias towards class 5, that is, amphitelic states are unstable.
Note that when $\alpha\ne0, \beta=0$, the process eventually ends up in either $r_n(n,0,0,n)$ or $r_n(0,n,n,0)$, 
that is, the class 5 states with maximal number of kMTs.

\section{Steady-state PMF (probability mass function) in Model II}

To calculate the steady-state PMF in Model II, consider it as a process of choosing the number of microtubules per kinetochore and 
distributing them to left and right poles. 
Let $k (\leq  n)$ be the total number of microtubules attached to the kinetochore and \(\phi _n(i,j)\) be the PMF for
state \(s_n(i,j)\), then 
$ \underset{i=0}{\overset{k}{\sum  }}\phi _n(i,j)=u_k, \;i+j=k.$
\noindent \(\phi _n(i,j)\) is derived by distributing \(u_k\) according to the binomial distribution: 
\begin{linenomath*}
\begin{equation*}
\binom{i+j}{i}
(1/2)^i(1/2)^j=
\binom{k}{i}(1/2)^k, \;0\leq  i\leq  k.
\end{equation*}
\end{linenomath*}
\noindent Hence, using Eq. (\ref{sol1}),
\begin{linenomath*}
\begin{eqnarray}
 \phi _n(i,j) &=& u_k\times 
\binom{k}{i}(1/2)^k \nonumber\\
 &=& \left(1+\frac{\rho }{n}\right)^{-n}\left(\frac{\rho }{2n}\right)^{i+j}\frac{n!}{i!j!(n-i-j)!}.
 \label{sol2}
\end{eqnarray}
\end{linenomath*}
\noindent Let \(\Phi _n=\left( \phi _n(0,0), \phi _n(0,1), \phi _n(1,0), \phi _n(0,2), \phi _n(1,1),\text{...},\phi _n(n-1,1) ,\phi _n(n,0) \right)\). 
Then, by applying Eq. (\ref{rule2}) and (\ref{sol2}), we find \(\Phi _n.P_n=\Phi _n\), which is consistent with the equilibrium at steady states.

\section{Size of the Markov chains}

The size of a Markov chain in Model II (total number of states) corresponds to the maximum of the \(s_n(i,j)\) indices according to 
Eq. (\ref{index}), which is \((n+1)(n+2)/2\).
The total number of states in the full model (Model III) is thus \((n+1)^2\left.(n+2)^2\right/4\), which grows rapidly as \(n\) increases
(Fig. S2A). 
Note that class 4 becomes predominant as the system size gets larger (Fig. S2B). 
Consequently, the number of possible state transitions also increases exponentially with the system size (Fig. S2C), 
which corresponds to the number of non-zero entries in the probability transition matrix.

\section{First passage time to class 5}

For a Markov chain of Model III, the mean first passage time \(f_i\) to class 5 from state\textit{  }\(i\) is obtained 
as the solution of linear equations \cite{bertsekas2008introduction}:
\begin{linenomath*}
\begin{equation*}
 f _i= \left\{
\begin{array}{ll}
 1+\underset{j\notin \text{class 5}}{\sum  }p_{i, j} f_{ j},  & i\notin \text{class 5}, \\
 0, & i\in \text{class 5}. 
\end{array}
\right.
\end{equation*}
\end{linenomath*}
\noindent \(f_1\) in Fig. 2A was calculated by incrementing \(p\) and \(q\) by 0.005 (\(50\times 10\) points) and 
in Figs. S3A and S3D by 0.0001 (\(100\times 100\) points). 
For a given value of \(q\) (\(=0.0005\)), the minimum \(f_1\) plateaus as \(n\) grows (Fig. S3B), 
although the number of states and transitions increase rapidly (Figs. S2A and S2C). 
For meiosis I, the \(q/p\) ratio for the minimum \(f_1\) approaches $\sim $1 as \(n\) increases (Fig. S3C).
For mitosis, the optimal \(q/p\) ratio is somewhat skewed (Fig. S3D).

\section{PMF time series of \boldmath\(r_n\left(i_1,j_1,i_2,j_2\right)\)}

It is straightforward to calculate the PMF of \(r_n\left(i_1,j_1,i_2,j_2\right)\) at each time point from the transition probability matrix. 
 We classified the PMFs according to Fig. 1D and calculated the sums for each class to obtain Fig. 2B to G. 
 Fig. S4 shows the PMF time series in meiosis I with $\alpha =0,\beta=1$ (a) and $\alpha =1,\beta=0$ (b) ($n=10, p=q=0.05$). 
 Class 4 is predominant in these condition as well. 

The dynamics are qualitatively very similar with any \(n\); 
this is mainly because the structure of the Markov chain remains the same as the size
of the chain grows (Fig. 1B and S2E). 
We have used \(n=10\) for most of the analysis as a representative value. 
We extensively explored the dynamics with different values of \(n\) and found fundamentally no diffence in the behavior of the Markov chain by altering \(n\) (for example, $n=10$ versus $15$ in Figs. S5B, C, E and F).

\section{Invariant dynamics of the Markov process with constant \boldmath${q/p}$ ratio}

As long as \(q/p\) ratio (relative kMT dissociation rate) remains the same, the steady-state probabilities of \(r_n\left(i_1,j_1,i_2,j_2\right)\)
stay the same for all \(p,q\) pairs. The PMF time series are also almost invariable (but with different time scales) as long as \(q/p\) ratio remains
constant, illustrated by the examples shown in Fig. S5. 
Only the time scale changes, which is inversely proportional to \(\sqrt{p \,q}\).
Strictly speaking, although steady-state probabilities are identical, PMFs at any given moment are not exactly the same: 
these small differences come about by the assumption that only a single event happens in every state transition. 
Differences are small enough to be ignored when \(p\) and \(q\) are sufficiently small. 
From a biological perspective, the change in time scale with the same dynamics has a different meaning: the faster
the association and dissociation of kMTs, the more efficient the chromosome biorientation. See also the section 'Biorientation attempts' below.

\section{Probability distribution of kMTs at steady states in meiosis I.}
\subsection*{Class 5}

\noindent Steady state probability distributions of the number of kMTs in class 5 for 
\(n=10, \text{  }p=0.1, \text{  }q=0.05\) are shown 
in Fig. S8B as density plots. 
Total probability of class 5 is indicated for each panel. 
Gray scale is normalized to the total probability of class 5. 
When $\alpha $ and $\beta $ are sufficiently small, class 5 (amphitelic) states are stable, 
i.e. the number of kMTs are close to the maximum. 
Otherwise, only a few microtubules on average are attached to each kinetochore.

\noindent
\subsection*{Class 1 to 4}

\noindent Steady state probability distributions of the number of kMTs in classes 1 to 4 for 
$n=10$, $p=0.1$, $q=0.05$ are shown 
in Fig. S8C as density plots. 
Although the total probabilities are greatly affected by the parameters $\alpha $ and $\beta $, the distribution of 
the number of kMTs of non-class 5 states barely changes. 
This is presumably because the size of the non-amphitelic classes (mainly class 4) in total is significantly larger than that
of class 5 (Fig. S2B), buffering the influence of class 5. 
Thus, $\bar{N}=n\rho/(n+\rho)$ (the exact solution in the random condition $\alpha=\beta=\gamma=1$) is also an approximate of 
the steady state distribution of kMTs in meiosis I for non-amphitelic states when $\alpha\ne1,0<\beta\ne1$.

\section{Number of kMTs at steady states in class 5\,---\,an analytical approximation}

When \(\alpha =0\), the number of kMTs for states in class 5 can be estimated analytically without explicitly calculating the PMF, 
which is computationally expensive for large \(n\). 
The reason why it is possible becomes apparent by looking at the Markov chain{'}s structure and transitions\,---\,when {
}\(\alpha =0\), transitions from class 5 to class 2 are still possible, but not to class 4 anymore. 
Note that for large \(n\) the number of transitions between class 5 and class 4 is far larger than
the one between class 5 and class 2 (Fig. S2D).

Because of its limited communication with other classes when \(\alpha =0\), 
class 5 behaves as if it is a disjoint class at steady states. 
Its two sub-classes (e.g. left and right square grids in Fig. S2E) have the identical probability distribution by symmetry, 
which can be approximated by
\begin{linenomath*}
\begin{eqnarray*}
\psi _n(i,j)\text{}&=&\phi _n(i,0)\times \phi _n(0,j) \\
&=& 2^{-(i+j)}u_i\,u_j  \\
&=&
\binom{n}{i}\binom{n}{j}
\left(\frac{\bar{\rho }}{2n}\right)^{i+j}\left(1+\frac{\bar{\rho }}{n}\right)^{-2n},
\end{eqnarray*}
\end{linenomath*}
\noindent where \(1\leq  i,j\leq  n \text{ and } \bar{\rho }=\rho /\beta =\frac{2p}{\beta  q}\). 
We now compute the conditional expectation of the number of kMTs \(i\) (or \(j\)) given the state \textit{s} is in class 5: 
\begin{linenomath*}
\begin{eqnarray*}
E(i|s\in \text{class 5}) &=& \underset{i=1}{\overset{n}{\sum  }}\,\underset{j=1}{\overset{n}{\sum  }}\left( \psi _n(i,j)\text{}\times i \right)\Big/\underset{i=1}{\overset{n}{\sum}}\,\underset{j=1}{\overset{n}{\sum  }}\psi _n(i,j)\text{}  \\
 &=&\bar{N}_5.
\end{eqnarray*}
\end{linenomath*}
\noindent After a lengthy algebra, it simplifies to Eq. (3) in the main text.
Similarly, 
\begin{linenomath*}
\begin{equation*}
E\left(\left.i^{ 2}\right|s\in \text{class 5}\right) = 
\underset{i=1}{\overset{n}{\sum  }}\,\underset{j=1}{\overset{n}{\sum  }}\left( \psi _n(i,j)\text{}\times i ^2\right)\Big/\underset{i=1}{\overset{n}{\sum
 }}\,\underset{j=1}{\overset{n}{\sum  }}\psi _n(i,j)\text{}. 
\end{equation*}
\end{linenomath*}
\noindent After another lengthy calculation, variance of \(i\) is reduced to:
\begin{linenomath*}
\begin{eqnarray}
 \text{Var}(i| s\in \text{class 5}) &=& E\left(\left.i^{ 2}\right|s\in \text{class 5}\right)-( E(i|s\in \text{class 5})\, )^2 \nonumber\\
&=& \frac{\bar{\rho } \left(2+\frac{\bar{\rho }}{n}\right)^{n-2} \left(2\left(2+\frac{\bar{\rho }}{n}\right)^n-2^n \left(2+\bar{\rho }\right)\right)}{\left(\left(2+\frac{\bar{\rho}}{n}\right)^n-2^n\right)^2}.
\label{eqnvar}
\end{eqnarray} 
\end{linenomath*}
\noindent Eqs. (3) and (\ref{eqnvar}) fit very well to the exact number of kMTs derived from the steady-state PMF 
(Fig. S8A). 
When \(n\) is small (e.g. \(n=4\)), the approximation diverges a little from the exact values, but is still pretty good (not shown). 
The mean and variance approach \textit{n} and 0, respectively,
as \(\bar{\rho }\to  \infty\).

\section{Probability of synchrony}

\subsection*{Computing the probability of synchrony}

\noindent We compute the probability of synchrony at time \(T=t\), i.e. the probability that the process in every chain is 
in the same class (in particular class 5) at the same time. 
This is illustrated by an example below, which shows the state (class) transtions of each process (Ch.1 to 4) from 
$T=0$ to $20$. 
Synchrony in class 5 is highlighted in red, which occurs at \(T=18\). \newline
\begin{linenomath*}
\noindent\(\begin{array}{r|rrrrrrrrrrrrrrrrrrrrr}
  &  0 & 1 &  2 & 3 &  4 &  5 & 6 &  7 &  8 &  9 & 10 & 11 & 12 & 13 & 14
& 15 & 16 & 17 & 18 & 19 & 20 \\
\hline
 \text{Ch.1} & 1 & 2 & 1 & 1 & 1 & 1 & 1 & 1 & 2 & 3 & 4 & 4 & 4 & 4 & 4 & 2 & 2 & 2 & {\color{red}5} & {\color{red}5} & {\color{red}5}
 \\
 \text{Ch.2} & 1 & 1 & 1 & 1 & 1 & 1 & 1 & 1 & 2 & 4 & 4 & 4 & 4 & 4 & 4 & 4 & 5 & 5 & {\color{red}5} & {\color{red}5} & {\color{red}5} \\
 \text{Ch.3} & 1 & 2 & 1 & 1 & 1 & 1 & 1 & 2 & 3 & 3 & 2 & 5 & 5 & 5 & 5 & 5 & 5 & 5 & {\color{red}5} & {\color{red}5} & {\color{red}5} \\
 \text{Ch.4} & 1 & 2 & 3 & 3 & 2 & 2 & 2 & 3 & 4 & 4 & 4 & 4 & 5 & 5 & 5 & 5 & 5 & 5 & {\color{red}5} & {\color{red}5} & {\color{red}5} \\
\end{array}\) 
\end{linenomath*}

\noindent Let \(s[t]\) be the state of a single Markov process at time \(T=t\), \(a_j^{(t)}\) be the probability of \(s[t]=j\) and \textit{ k} be the total
number of chains. In the above example, \(k=4\). Also let \(P_s^{(t)}\) and \(P_{\text{\textit{as}}}^{(t)}\) be the probability of synchrony and
asynchrony at \(T=t\), respectively. Then,
\begin{linenomath*}
\begin{eqnarray}
P_s^{(t)} &=& \theta _t{}^k, \nonumber\\
 P_{\text{\textit{as}}}^{(t)} &=& 1-P_s^{(t)}, \nonumber\\
 &=&1-\theta _t{}^k, \label{async}
\end{eqnarray}
\end{linenomath*}
\noindent where 
\begin{linenomath*}
\begin{equation*}
 \theta _t=P(s[t]\in \text{class 5}) 
 =\underset{j\text{   }}{\sum  }a_j^{(t)},\text{   }j\in \text{class 5} . \\
\end{equation*}
\end{linenomath*}

\noindent We used Eq. (\ref{async}) for Figs. 4A and S3E. 
Now we consider the probability of synchrony attempts. 
For this, we need some events and their probabilities defined.
The probability of biorientation attempts \(P_+^{(t)}\) in a single process is
\begin{linenomath*}
\begin{eqnarray*}
 P_+^{(t)}&=&P(s[t-1]\notin \text{class 5} \land  s[t]\in \text{class 5}) \\
& =&\underset{i, j}{\sum  }a_i^{(t-1)}p_{i, j},\text{   }i\notin \text{class 5}, \text{   }j\in \text{class 5},
\end{eqnarray*}
\end{linenomath*}

\noindent where \(p_{i, j}\) is the transition probability from state \(i\) to \(j\). 
Likewise, the probability of biorientation loss \(P_-^{(t)}\) is 
\begin{linenomath*}
\begin{equation*}
P_-^{(t)} 
=\underset{i, j}{\sum  }a_j^{(t-1)}p_{j, i},\text{   }i\notin \text{class 5}, \text{  } j\in \text{class 5}.
\end{equation*}
\end{linenomath*}
The probability of biorientation maintainance \(P_0^{(t)}\) of a process is
\begin{linenomath*}
\begin{eqnarray*}
P_0^{(t)} &=& P(s[t-1]\in \text{class 5} \land  s[t]\in \text{class 5}) \\
 &=& \underset{i, j}{\sum  }a_i^{(t-1)}p_{i, j},\text{   }i\in \text{class 5}, j\in \text{class 5}.
\end{eqnarray*}
\end{linenomath*}
\noindent Let \textit{ m} be the number of Markov processes in class 5 at \(T=t\). 
Note that a synchrony attempt occurs only when all \(m\) processes that
are in class 5 stay in class 5 and the remaining \(k-m\) processes undergo transition from non-class 5 to class 5 states. 
The probability of such a synchrony attempt, \(P_{\text{as}, s}^{(t)}\), is
\begin{linenomath*}
\begin{eqnarray*}
P_{\text{\textit{as}}, s}^{(t)} &=& P(\text{asynchrony } \text{at }T=t-1\land \text{synchrony } \text{at } T=t) \\
 &=& \underset{m=0}{\overset{k-1}{\sum  }}
\binom{k}{m}
\left(P_+^{(t)}\right){}^{ k-m}\left(P_0^{(t) }\right){}^m,
\end{eqnarray*}
\end{linenomath*}
\noindent where $\binom{k}{m}$ is the binomial coefficient. 
 Note that Fig. S9A may help understand the following derivation of formulae.\newline

\noindent The probability of synchrony maintenance \(P_{s, s}^{(t)}\) is
\begin{linenomath*}
\begin{eqnarray*}
P_{\text{\textit{$s$}}, \text{\textit{$s$}}}^{(t)} &=& P(\text{synchrony } \text{at } T=t-1\land \text{synchrony } \text{at } T=t) \\
&=& P_s^{(t)}-P_{\text{\textit{as}}, s}^{(t)}.
\end{eqnarray*}
\end{linenomath*}
\noindent Likewise, the probability of asynchrony maintenance \(P_{\text{\textit{as}}, \text{\textit{as}}}^{(t)}\) is
\begin{linenomath*}
\begin{eqnarray*}
P_{\text{\textit{as}}, a\text{\textit{$s$}}}^{(t)} &=& P(\text{asynchrony } \text{at } T=t-1\land \text{asynchrony } \text{at } T=t) \\
&=& P_{\text{\textit{as}}}^{(t)}-P_{\text{\textit{$s$}}, \text{\textit{as}}}^{(t)}.
\end{eqnarray*}
\end{linenomath*}
\noindent The probability of synchrony loss \(P_{s, \text{\textit{as}}}^{(t)}\) is obtained by
\begin{linenomath*}
\begin{eqnarray}
 P_{s, \text{\textit{as}}}^{(t)} &=& P(\text{synchrony } \text{at } T=t-1\land \text{asynchrony } \text{at } T=t) \nonumber\\
 &=& P_s^{(t-1)}-P_{s,s}^{(t)} \nonumber\\
 &=& P_s^{(t-1)}-\left(P_s^{(t)}-P_{\text{\textit{as}}, s}^{(t)}\right) \nonumber\\
 &=& P_{\text{\textit{as}}, s}^{(t)}-\left(P_s^{(t)}-P_s^{(t-1)}\right).
\end{eqnarray}
\end{linenomath*}
\noindent $P_{s, as}^{(t)}$ can also be obtained by 
\begin{linenomath*}
\begin{equation*}
P_{s, \text{\textit{as}}}^{(t)}=\underset{m=1}{\overset{k}{\sum  }}
\binom{k}{m}
\left(P_-^{(t)}\right){}^{ m}\left(P_0^{(t) }\right){}^{k-m}.
\end{equation*}
\end{linenomath*}
\noindent At steady states, the conditional probability of synchrony loss given the present state is in synchrony is 
\(P_{\text{\textit{$s$}}, \text{\textit{as}}}^{(\infty)}/P_s^{(\infty )}\), therefore the mean duration (half-life) of synchrony is 
\(P_s^{(\infty )}/P_{\text{\textit{$s$}}, \text{\textit{as}}}^{(\infty)}\). Fig. 4C was derived by this formula.

Now we examine when the synchrony happens for the first time, i.e. the probability of the first synchrony at \(T=t\), 
denoted by \(P_{\text{\textit{fs}}}^{(t)}\).
With \(\alpha =\beta =0\), once a process is in class 5, it is trapped in the class (i.e. \(P_{s, \text{\textit{as}}}^{(t)}=0\)). 
Therefore \(P_{\text{\textit{fs}}}^{(t)}=P_{\text{\textit{as}}, s}^{(t)}\). 
When $\alpha $ and $\beta $ are small, the majority of synchrony attempts are for the first time; 
in addition, as $k$ gets larger (number of processes, i.e. pairs of sister chromatids in mitosis or bivalents in meiosis I), 
synchrony becomes a rarer event. 
Thus, the probability of synchrony loss $P_{s, \text{\textit{as}}}^{(t)}$ are small at any given moment for large $k$ and 
small $\alpha $ and $\beta $. 
In such a condition, it is therefore possible to approximate 
$P_{\text{\textit{fs}}}^{(t)}$ with $\tilde{P}_{\text{\textit{fs}}}^{(t)}$:
\begin{linenomath*}
\begin{eqnarray*}
\tilde{P}_{\text{\textit{fs}}}^{(t)}
&=& P_{\text{\textit{as}}}^{(0)} \\
& &\times
\left(\underset{\tau =1}{\overset{t-1}{\prod  }}P(\text{asynchrony } \text{at } T=\tau -1\land  \text{asynchrony }
\text{at } T=\tau |\text{asynchrony } \text{at } T=\tau -1)\right) \\
& &\times
 P(\text{asynchrony } \text{at } T=t-1\land  \text{synchrony } \text{at } T=t|\text{asynchrony }
\text{at } T=t-1) \\
 &=& \left(\underset{\tau =1}{\overset{t-1}{\prod  }}\frac{P_{\text{\textit{as}}, \text{\textit{as}}}^{(\tau )}}{P_{\text{\textit{as}}}^{(\tau -1)}}\right)\times
\frac{P_{\text{\textit{as}}, s}^{(t)}}{P_{\text{\textit{as}}}^{(t-1)}} , \text{  } t=2,3,4\text{...}\,. 
\end{eqnarray*}
\end{linenomath*}
\noindent It is apparent that for $\alpha =\beta =0$ (therefore $P_{s, \text{\textit{as}}}^{(t)}=0$), 
$ \tilde{P}_{\text{\textit{fs}}}^{(t)} =P_{\text{\textit{as}}, s}^{(t)}$. 
\noindent Fig. S9B shows an example of $\tilde{P}_{\text{\textit{fs}}}^{(t)}$ together with Monte Carlo simulation results 
(probability in 5,000 simulations), demonstrating a good fit of the approximation to the simulation result.

\subsection*{Timing of first synchrony and $q/p$ ratio}

\noindent We asked how $q/p$ ratio affects the timing of first synchrony. 
We also asked how efficiently synchrony can be achieved in a slightly compromized
condition, i.e. \(\alpha =\beta =0.05\). 
Fig. S9C shows the probability of first synchrony in meiosis I at each time point with decreasing 
\(q\) value (\(n=10,\text{ }p=0.05\)
and the number of bivalents \(k=5)\). 
When \(p=q=0.05\), first synchrony happens most frequently around \(T=100\); 
By \(T=400\) synchrony takes place at least once in \(\sim 99.7\%\) of cases (not shown). 
As \(q/p\) ratio declines, the timing of first synchrony spreads more and more over time, becoming unpredictable. 
Therefore, synchrony does happen relatively efficiently with the right \(q/p\) ratio even in a slightly compromized condition 
with $\alpha =\beta =0.05$.
For a fixed value of \(p=0.05\), the probability of synchrony at steady state (at any give moment) is 0.66 with \(q=0.05\), but only 0.017 
when \(q=0.01\).
For \(k=20\), the probability declines to 0.19 with \(q=0.05\) and \(8.3\times\)\(10^{-8}\) with \(q=0.01\). 
Thus, keeping the balance of \(q/p\)ratio is all the more important for the cell with a large number of chromosomes.
This principle applies to both mitosis and meiosis I.

\section{Bi-orientation attempts}

\subsection*{Probability of bi-orientation attempts}
\noindent Probability of biorientation attempts at time \(T=t\)\textit{ , }\(\mu _t\), is  
\begin{linenomath*}
\begin{equation*}
 \mu _t = \underset{i, j}{\sum  } a_i^{(t)}p_{i, j}, \;i\notin \text{class 5}, j\in \text{class 5},
\end{equation*}
\end{linenomath*}
\noindent where \(a_i^{(t)}\) is the probability of the process in state \(i\) at time $t$ and \(p_{i, j}\) is the transition probability 
from state \(i\) { }to \(j\). 
\(\mu _t\) can also be interpreted as the mean number of attempts to biorientation at time $t$. 
Fig. S10A shows a plot of \(\mu _t\) by this formula (analytical solution) and simulations 
(parameters: \(n=5, p=q=0.01, \alpha =\beta =0.1\); 10,000 simulations).
 Fig. S10B shows an example of probability time series of biorientation attempts, with \(p=q=0.05\) versus 
\(p=q=0.01\) ($n=10$, $\alpha =\beta =0.1$).
With the same \(q/p\) ratio, their PMF time series are almost identical (not shown) if the time scale is adjusted. 
Because of this change of time scale, the probability (i.e. frequency) of biorientation attempts also changes. 
In this example, the probability at steady states (at any time point)
decreases from $\sim $0.033 with \(p=q=0.05\) to $\sim $0.0066 with \(p=q=0.01\).

\subsection*{Mean number of biorientation attempts before absorption}

\noindent The number of biorientation attempts before the onset of anaphase is equivalent in our model 
to the total number of transitions to class 5 from either class 2 or class 4 before absorption 
(referred as $\bar{M}$ hereafter; Fig. 4D). 
This can be computed by first calculating the mean total number of times
the process is in each transient state before absorption, starting from class 1. 
We denote this number as $M(i)$, \(i\in (\text{transient } \text{states})\).
There are two absorbing states when \(\beta =0\), so the total number of transient states is 
\(l=(\text{total } \text{number } \text{of } \text{states})-2\).
{ }\(M(i)\) is obtained from the so-called fundamental matrix \(N\) defined as \(N=(I-Q)^{-1}\), where \(I\) is the \(l\times l\) identity matrix
and \(Q\) is the \(l\times l\) submatrix of \(P\) (the transition probability matrix):
\begin{linenomath*}
\begin{equation*}
P=\left(
\begin{array}{cc}
 Q & R \\
 O & S \\
\end{array}
\right).
\end{equation*}
\end{linenomath*}
\noindent \(Q\) defines the transition within the transient states. \(O\) is a \(2\times l\) matrix with all 0{'}s; 
\(R\) concerns the transtion from transient to absorbing states; 
\(S\) is the \(2\times 2\) identity matrix in our model. 
The first row of \(N\) corresponds to \(M(i)\). 
The formula for fundamental matrices (and also for the mean and variances of absorption time) is 
according to Kemeny and Snell \cite{Kemeny:1960gs}.
Computing the mean of \(\bar{M}, \left\langle \bar{M}\right\rangle\), is straightforward using \(M(i)\): 
\begin{linenomath*}
\begin{equation*}
 \left\langle \bar{M}\right\rangle = 
\underset{i}{\sum  }\,\underset{j}{\sum  }M(i)\,p_{i, j},
\end{equation*} 
\end{linenomath*}
\noindent where \(i\in (\text{class} 2\lor \text{class} 4)\), \(j\in \text{class 5}\) and \(p_{i, j}\) is the transition probability from state \(i\) to \(j\).
Note that, when \(\alpha =0,\bar{M}=1\) for all \(n\geq 1\) because once in class 5 the process never leaves the class.

\newpage

  \begin{figure}[h!]
  \includegraphics[width=12cm]{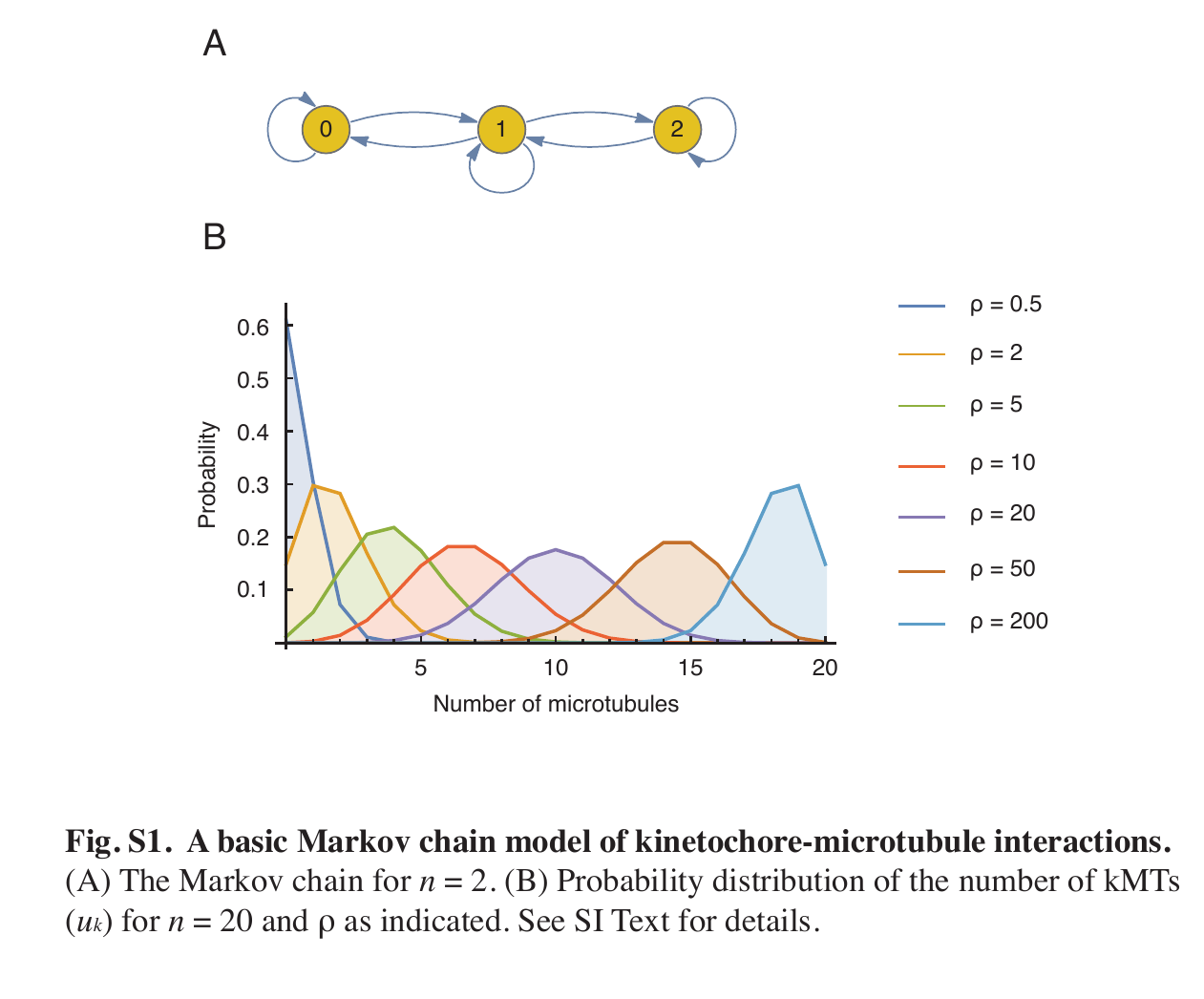}
       \end{figure}

\newpage

  \begin{figure}[h!]
  \includegraphics[width=16cm]{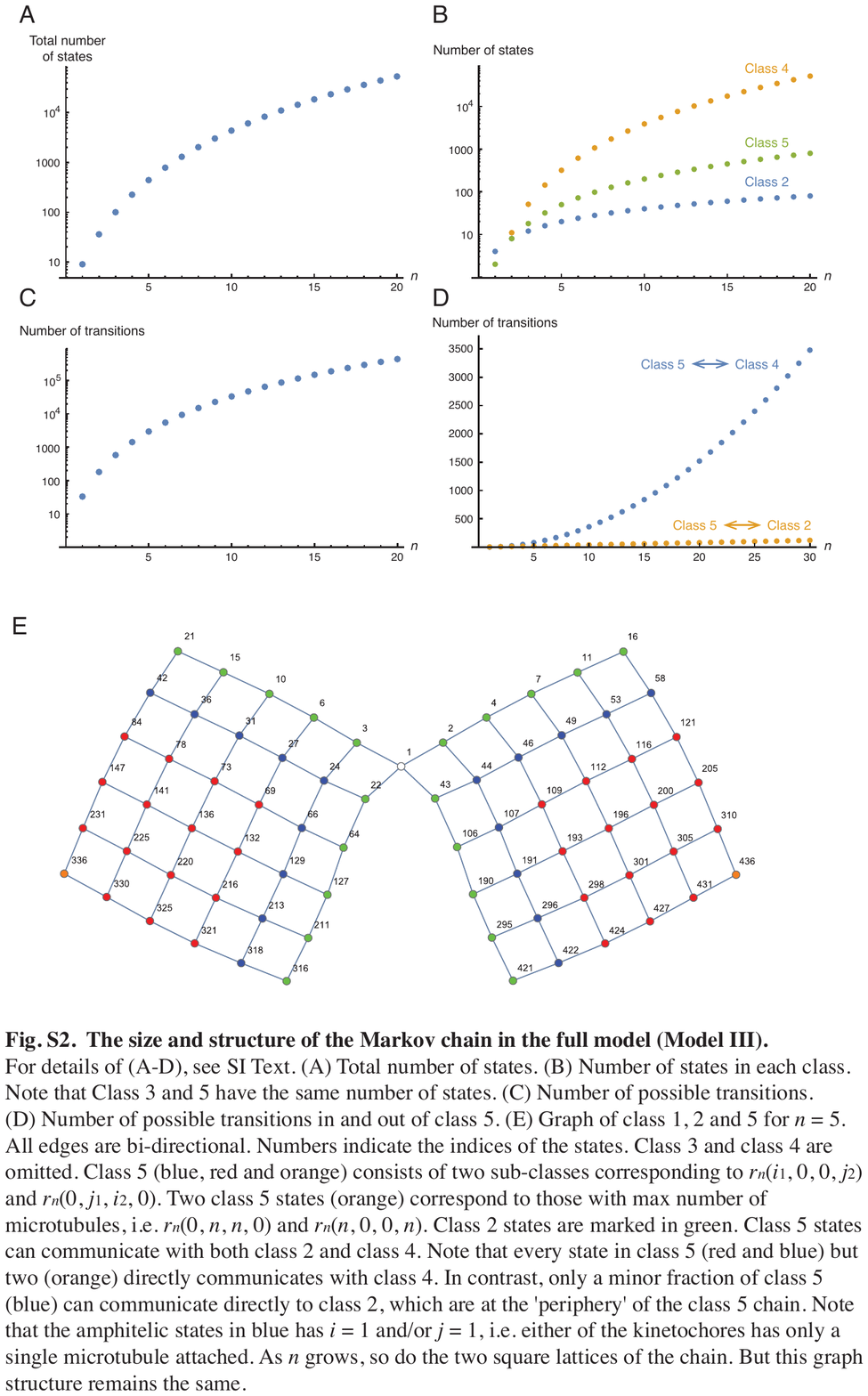}
       \end{figure}

\newpage

  \begin{figure}[h!]
  \includegraphics[width=15cm]{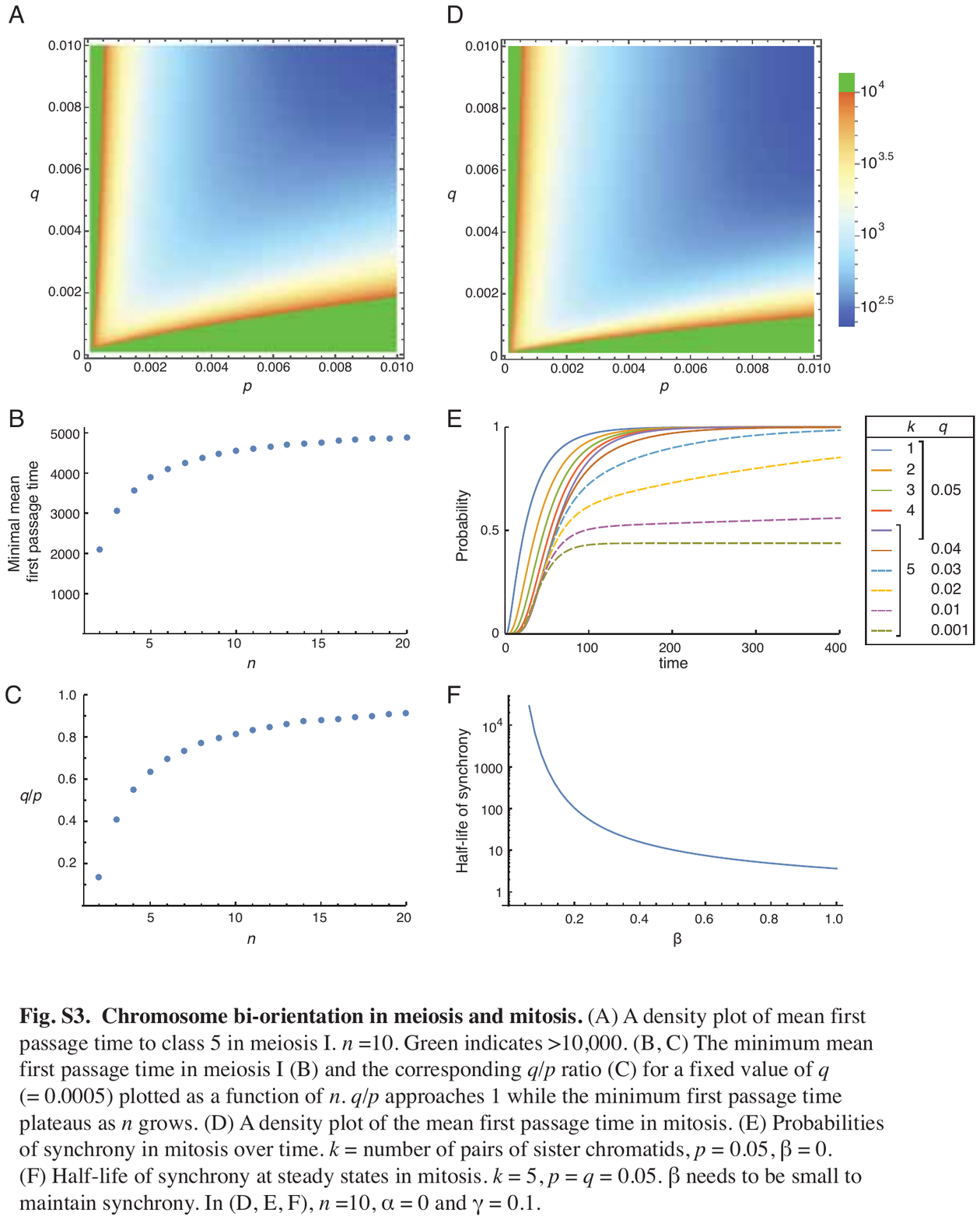}
       \end{figure}

\newpage

  \begin{figure}[h!]
  \includegraphics[width=12cm]{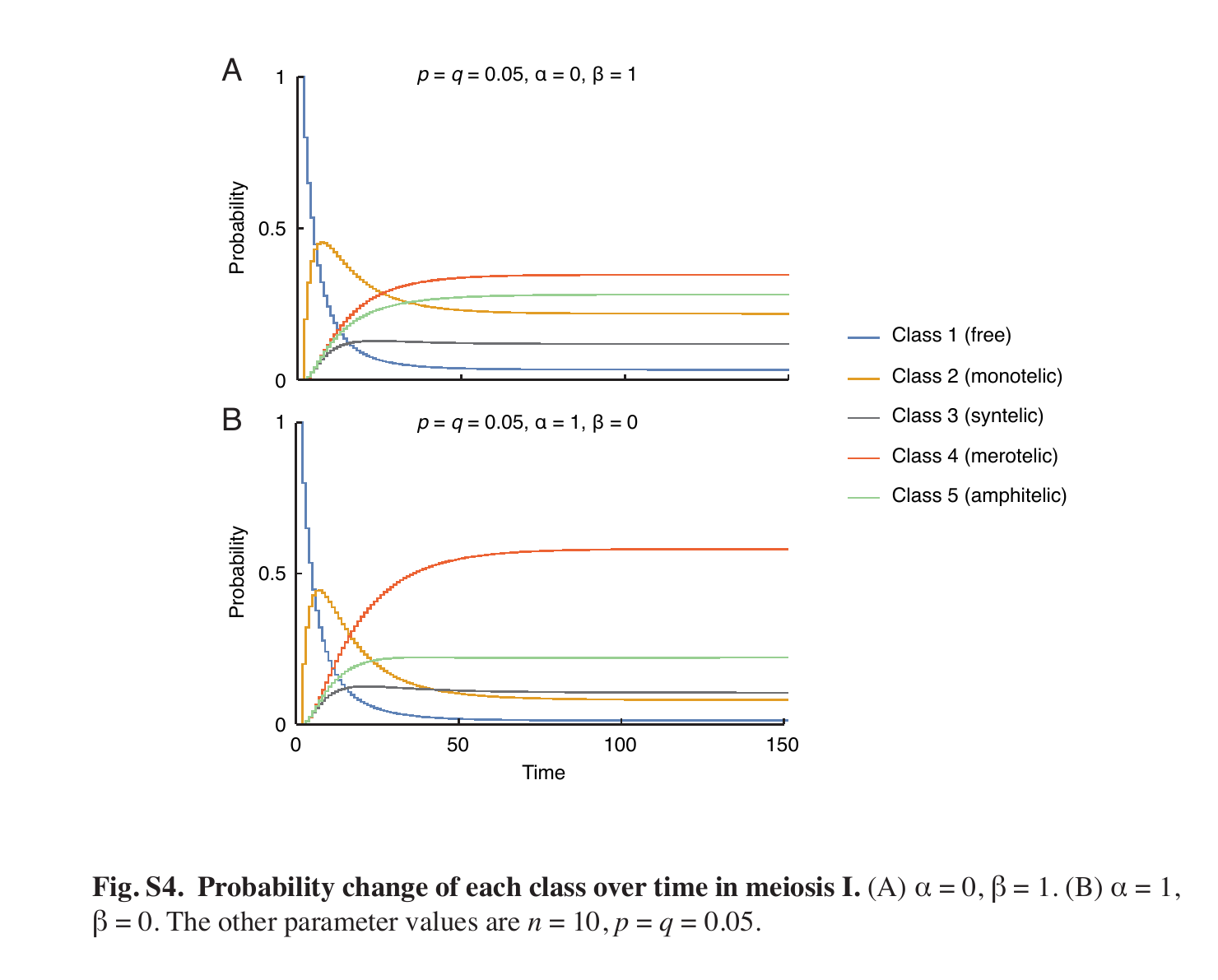}
       \end{figure}

\newpage

  \begin{figure}[h!]
  \includegraphics[width=16cm]{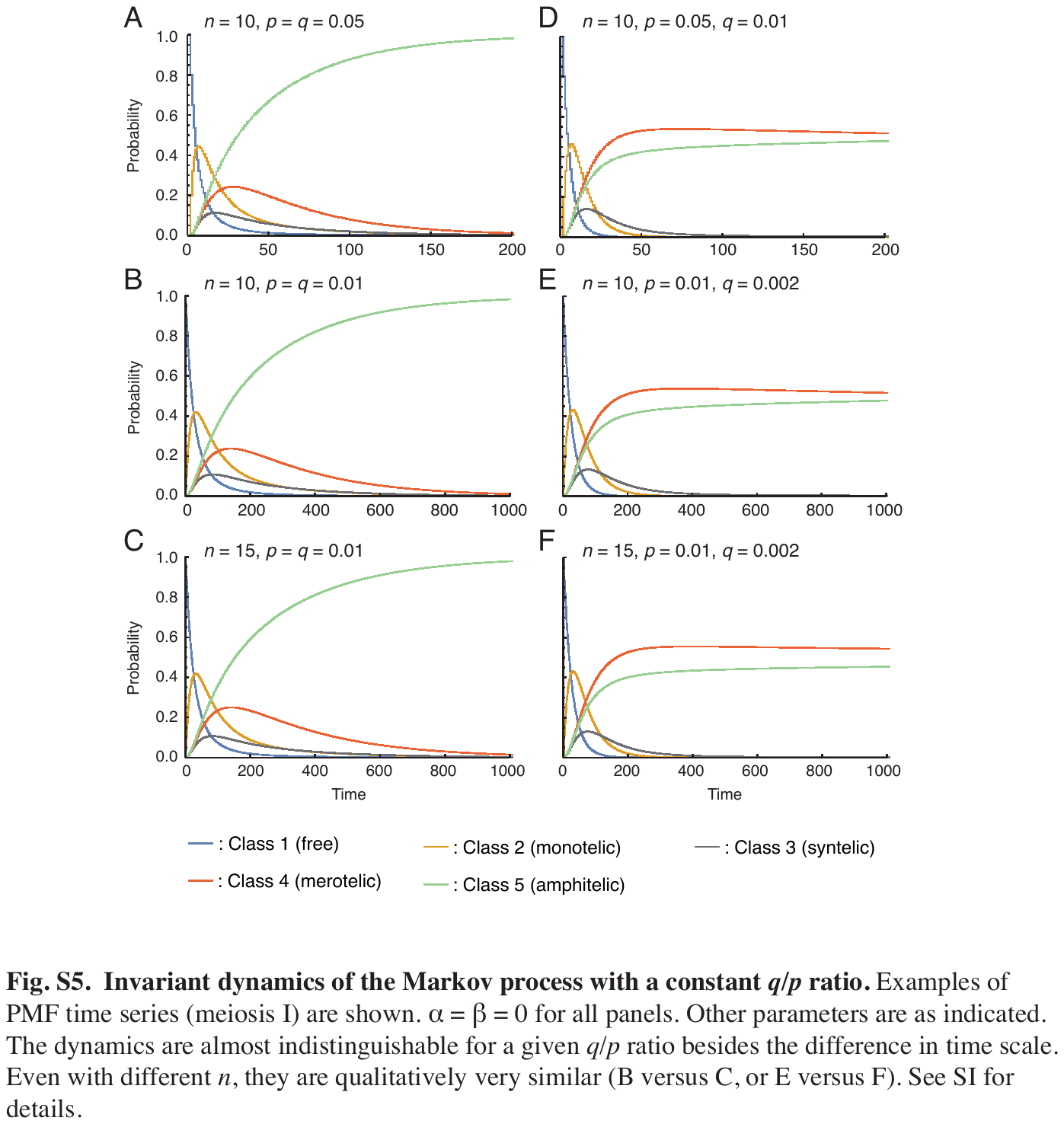}
       \end{figure}

\newpage

  \begin{figure}[h!]
  \includegraphics[width=16cm]{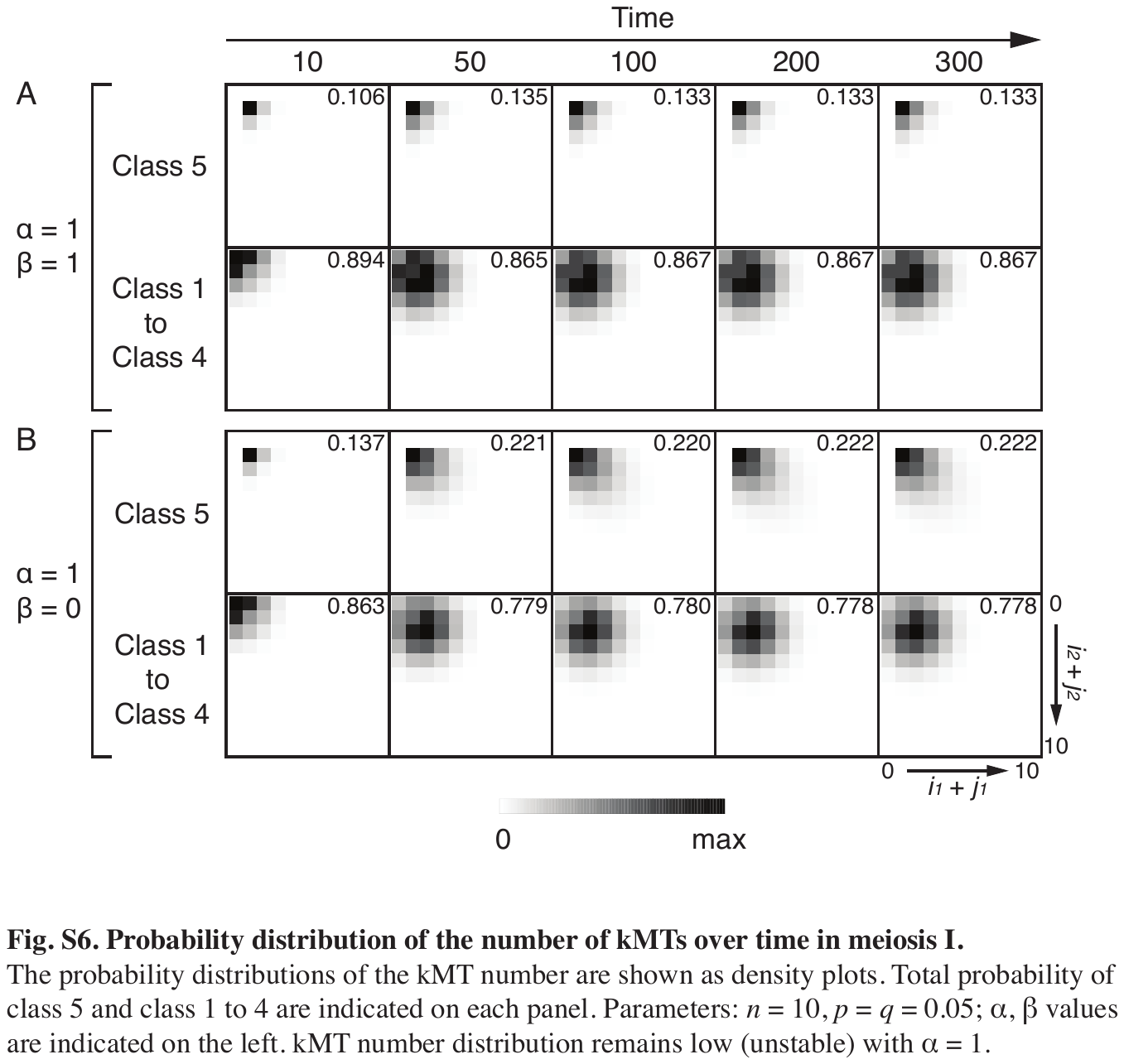}
       \end{figure}

\newpage

  \begin{figure}[h!]
  \includegraphics[width=16cm]{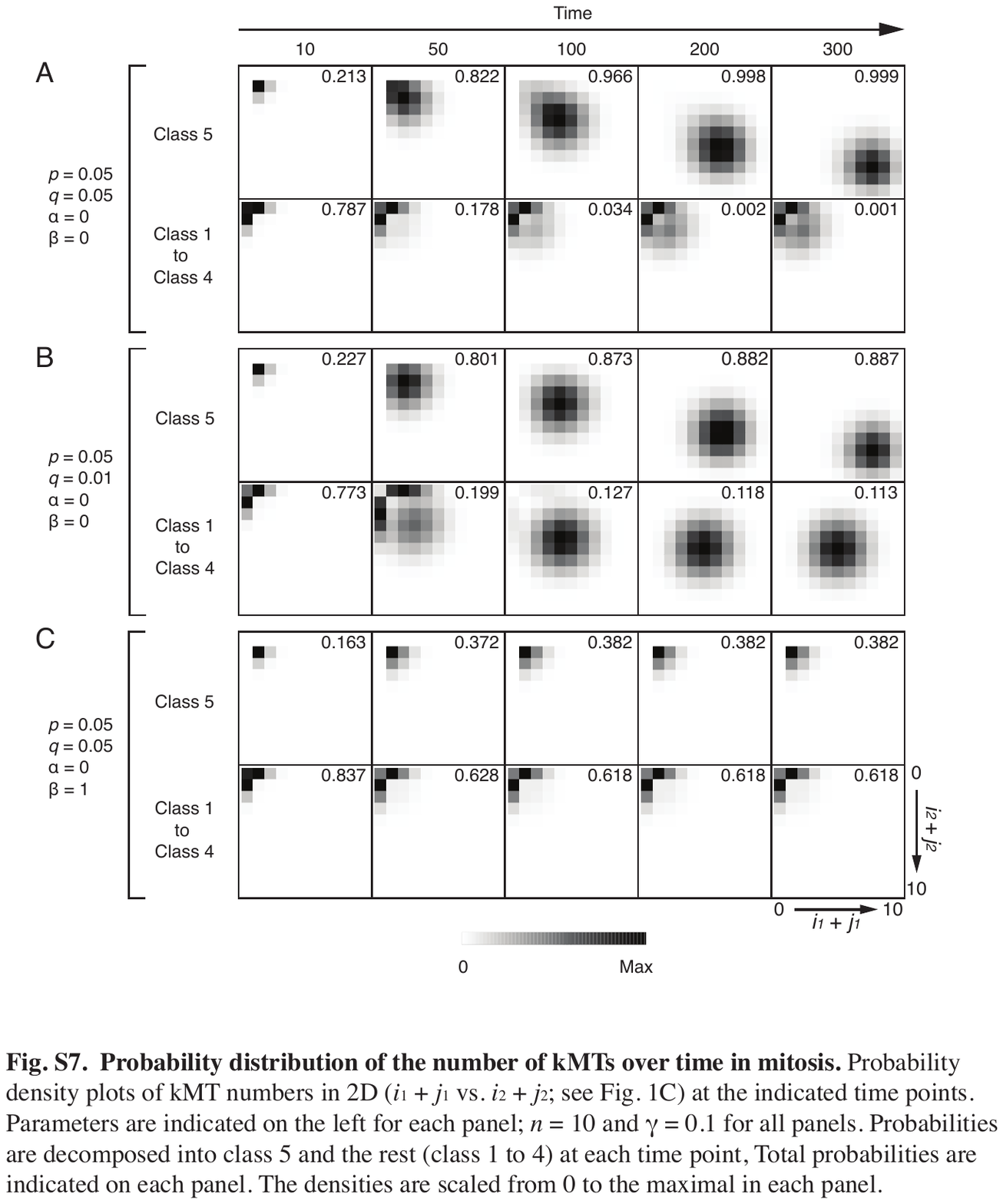}
       \end{figure}

\newpage

  \begin{figure}[h!]
  \includegraphics[width=16cm]{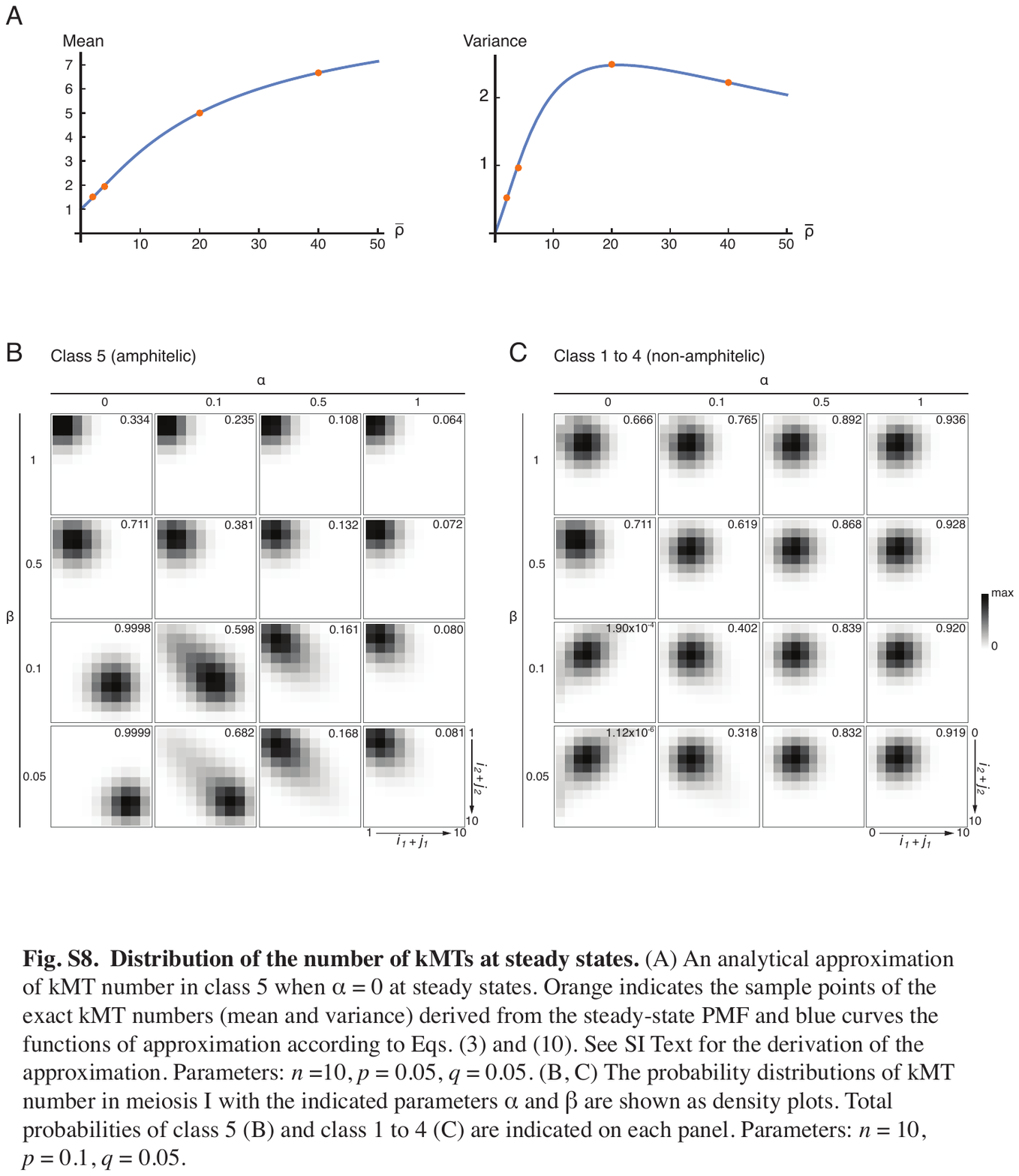}
       \end{figure}

\newpage

  \begin{figure}[h!]
  \includegraphics[width=16cm]{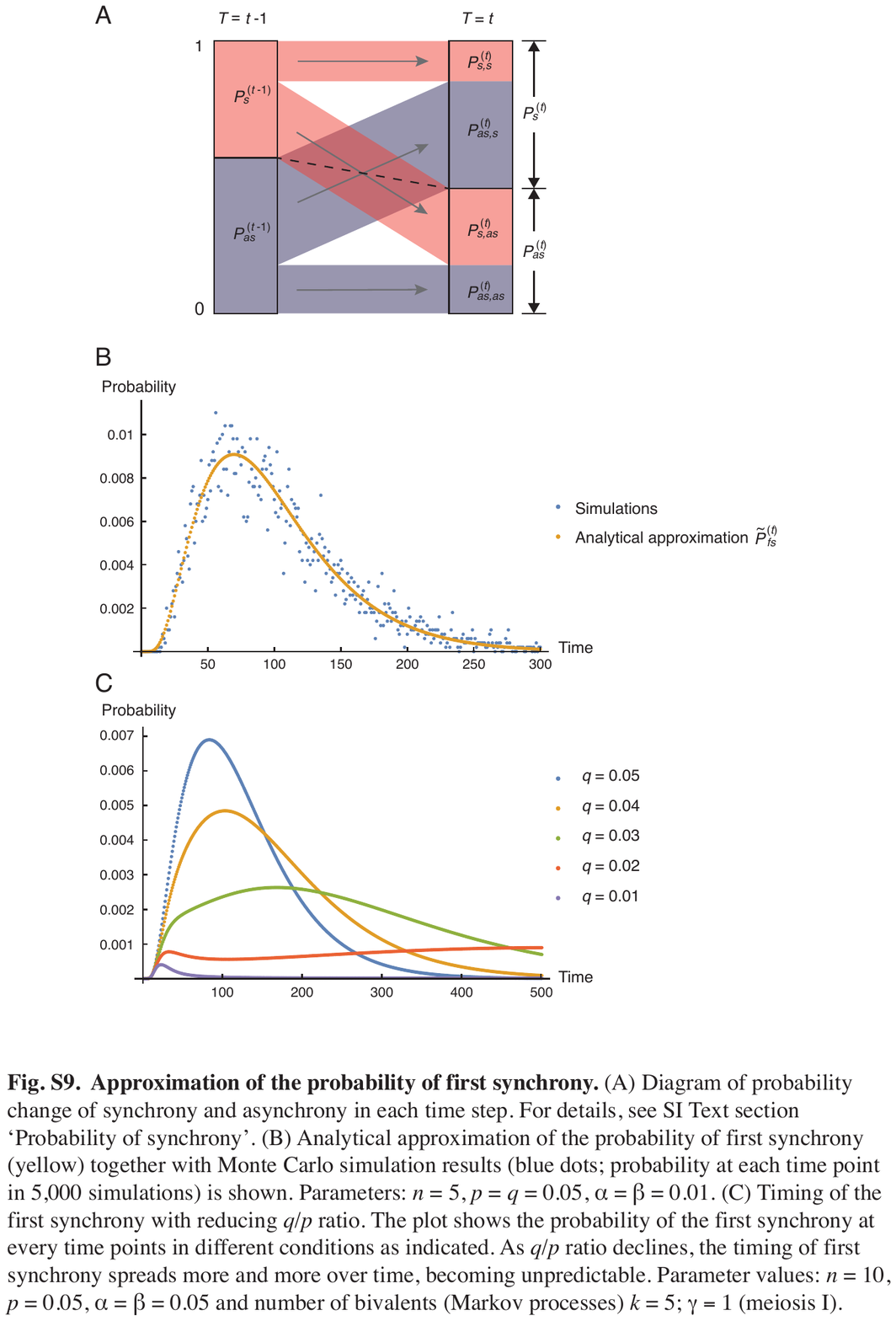}
       \end{figure}

\newpage

  \begin{figure}[h!]
  \includegraphics[width=16cm]{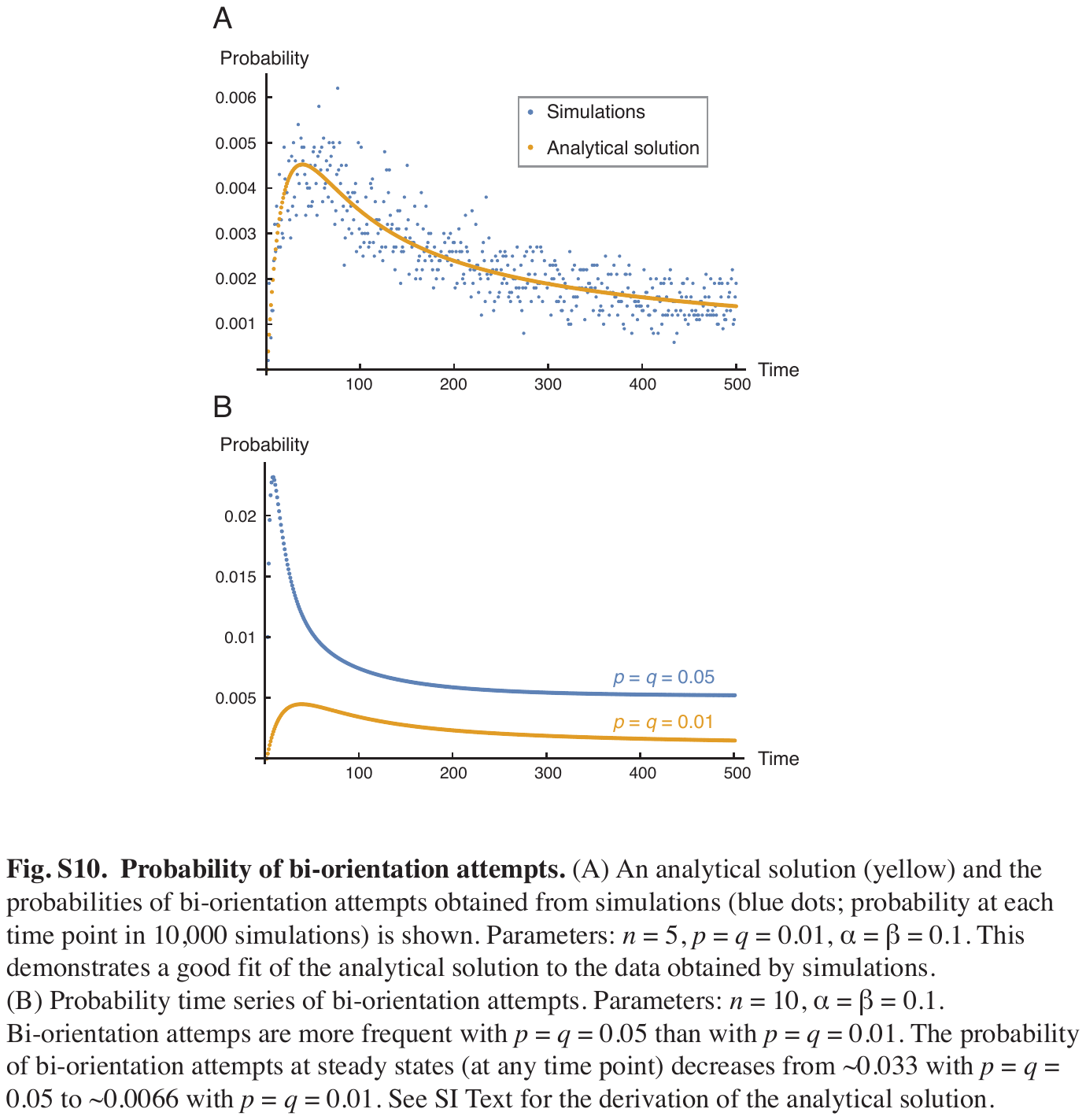}
       \end{figure}

\newpage

\end{document}